\documentclass[a4paper,11pt]{article}
\pdfoutput=1 

\usepackage{jheppub} 

\usepackage[T1]{fontenc} 
\usepackage{amsmath}
\usepackage{amsfonts}
\usepackage{amssymb}
\usepackage{appendix}
\usepackage{color}
\usepackage{physics}
\usepackage{dsfont}
\usepackage{mathtools}
\usepackage{subcaption}

\newcommand{\be}{\begin{equation}}
\newcommand{\ee}{\end{equation}}
\newcommand{\ba}{\begin{eqnarray}}
\newcommand{\ea}{\end{eqnarray}}

\title{\boldmath Scalar and fermionic Unruh Otto engines}

\author[a,b]{Finnian Gray}
\author[a,b,1]{and Robert B. Mann.\note{Corresponding author.}}

\affiliation[a]{Perimeter Institute,\\31 Caroline St. N., Waterloo, Ontario, N2L 2Y5, Canada}
\affiliation[b]{Department of Physics and Astronomy,\\ University of Waterloo, Waterloo, Ontario,  N2L 3G1, Canada}

\emailAdd{fgray@perimeterinstitute.ca}
\emailAdd{rbmann@uwaterloo.ca}

\abstract{We investigate the behaviour of quantum heat engines, in which a qubit is put through the quantum equivalent of the Otto cycle and the heat reservoirs are due to the Unruh effect.  The qubit is described by an Unruh--DeWitt detector model coupled  quadratically to a scalar field and alternately to a fermion field. In the cycle, the qubit undergoes two stages of differing constant acceleration corresponding to thermal contact with a hot and cold reservoir. Explicit conditions are derived on the accelerations required for this cycle to have positive work output. By analytically calculating the detector response functions, we show that the dimensionality of the quadratic and fermionic coupling constants introduces qualitatively different behaviour of the work output from this cycle as compared to the case in which the qubit linearly couples to a scalar field.
}

\begin{document}
\maketitle
\flushbottom

\section{Introduction}
 
The relationship between quantum physics and thermodynamics, a subject known as quantum thermodynamics, has in recent years developed into an active and lively field.  One topic of growing interest is that of using quantum systems to push the limits of classical heat engines~\cite{Maruyama2009}. A quantum version of the Otto cycle
has been introduced \cite{Kieu2004,Kieu2006}, in which a two level system (a qubit) is put through a series of steps mirroring the adiabatic and ischoric processes of the classical combustion engine.

A considerably different perspective comes out of gravitational physics. Since the 70s there has been deep insight in understanding the thermodynamic connection with black holes and gravity. In particular has been demonstrated that black holes obey a set of thermodynamic laws, with area playing the role of entropy~\cite{Bekenstein1972,Bardeen1973}.  
This phenomenon was thought to be little more than a curious coincidence until the connection was made concrete when Hawking showed that black holes radiate with temperature proportional to the surface gravity~\cite{Hawking75HR,Hawking76HR}. This can be understood from techniques of quantum field theory in curved spacetime, where one no longer has Poincar\'e symmetry and no unique vacuum~\cite{Birrell1984quantum}. 

Furthermore it has been demonstrated this thermality is predicted to occur in flat spacetimes for an accelerated observer. This is the celebrated Unruh effect~\cite{Unruh1976},  which makes manifest the connection between acceleration in relativistic settings and temperature of the thermal vacuum.  Different observers will see different numbers of particles. 

It is therefore necessary to introduce an operational concept of a particle.  This is given by the Unruh--DeWitt detector model~\cite{Unruh1976,DeWitt79}, wherein we measure a particle as an excitation of a quantum system via an interaction $\mathcal{H}_I\propto m\mathcal{O}_F$. Here $m$ is the monopole operator  describing the (de-)excitation of the detector, typically taken to be a two level system. $\mathcal{O}_F$ represents an operator describing the interaction of the detector with a quantum field. The standard model is simply a linear coupling to a scalar field, $\mathcal{O}_\phi=\phi$, but generalizations to higher spin~\cite{Takagi1986} and quadratic coupling~\cite{Hummer2016,Sachs2017} have been considered. The response function of this detector $\mathcal{F}(\Omega,\mathcal{T})$ represents the probability of detection over a time scale $\mathcal{T}$. For a constantly accelerated observer at large times, this response function approaches a thermal state in the rigorous Kubo--Martin--Schwinger (KMS) sense~\cite{Fewster2016,Garay2016}, with Unruh temperature proportional to the acceleration, i.e $T_U\propto\alpha$.
In addition, these Unruh--DeWitt (UdW) detectors have been used to study unusual phenomena, such as harvesting entanglement from the quantum vacuum~\cite{Reznik2003,Reznik2004,Salton2015}. 

However the connection between quantum thermodynamics and relativistic quantum systems in this setting is a newly developing arena. 
Recently  Arias et al.~\cite{Arias2018} proposed  that a two level system (qubit) undergoing acceleration can extract work from the vacuum in a variant of the thermodynamic Otto cycle~\cite{Kieu2004}. Using this set up, the accelerated qubit exploits the quantum vacuum as an approximately thermal bath via the Unruh effect. 

The utility of the linearly coupled model has, as noted above, led to the question as to how other field couplings 
 such as quantized spinor or vector fields, modify the effect. These coupling are more typical in nature than the linear scalar coupling, and are of fundamental interest in their own right.  In this paper we consider two such couplings in the context of the Unruh quantum Otto Engine: a quadratic scalar coupling and a coupling to fermionic fields.  We shall employ the machinery previously developed for these couplings~\cite{Hummer2016,Sachs2017,Louko2016Fermions} to see how the basic properties of the engine are modified.
 We find that  the dimensionality of the coupling constants of the UdW models  plays an important role, leading to qualitatively different features in the work output of the cycle.

We begin by  discussing in detail the different UdW models we consider in Section~\ref{sec:UdWmod} and illustrate how a uniformly accelerated sees a thermal vacuum. The effects of the different finite time interactions on the qubit's state are calculated in terms of the detectors' response function. Then, in Section~\ref{sec:UQOC}, we describe the thermodynamic and kinematic processes of the Unruh quantum Otto engine introduced by Arias et al.~\cite{Arias2018}. In Section~\ref{sec:Results} we present our results for the thermodynamic quantities of the cycle and compare the thermal output of the different interactions. Finally, in Section~\ref{sec:Conc} we summarise our results and present an outlook on further possible research. Throughout we use mostly minus signature metric $\eta_{\mu\nu}=\text{diag}(1,-1,-1,-1)$ and natural units such that the physical constants $\hbar=c=k_B=1$.

\section{Unruh--DeWitt detector models}\label{sec:UdWmod}
We consider a qubit coupled to a scalar and a fermionic field with generic point-like Unruh--DeWitt detector model. Since the fermionic coupling naturally is quadratic in the field~\cite{Takagi1986,Hummer2016,Sachs2017} we also consider a quadratic coupling to a scalar field in order to elucidate and distinguish  features arising 
from fermions and features arising from scalars.

The Hamiltonian of this system reads
	\begin{equation}\label{eq:FullH}
	\mathcal{H}=\mathcal{H}_0+\mathcal{H}_{\text{Int}} \qquad  \mathcal{H}_0=H_{\text{F}}+H 
	\end{equation}
where $H_\text{F}$ is the Hamiltonian of the scalar or fermionic field (see Appendix~\ref{App:QFT Con} for our conventions) and $H=\Omega \ket{e}\bra{e}$ is that of the two level system/qubit with ground state $\ket{g}$, excited state $\ket{e}$ and energy gap $\Omega$. Their interaction is that of an Unruh-DeWitt coupling along a worldline $x^\mu(\tau)=(t(\tau),\mathbf{x}(\tau))$
	\begin{equation}\label{eq:IntH}
	\mathcal{H}_\text{Int}=\lambda_F\;
	\chi_\mathcal{T}(\tau)m(\tau)\;\mathcal{O}_\text{F}[x(\tau)]\;,
	\end{equation}
where $\lambda_F$ is the strength of the interaction and where the switching function $\chi_\mathcal{T}(\tau)$ describes how the detector is switched on. 

The switching function plays an important role.  It is known that when $\chi_\mathcal{T}(\tau)$ is sharp, e.g. a step function, problems with finite time Unruh--DeWitt detectors emerge~\cite{Satz2007,Louko2006}. These are characteristic of finite time interaction detectors~\cite{Svaiter1992,Padmanabhan1996} and the divergences get worse in higher dimensions~\cite{Hodgkinson2011}. To avoid these artefacts, we will take $\chi_\mathcal{T}(\tau)$ to be smooth with a characteristic width $\mathcal{T}$ representing how long the detector is on. Now $m(\tau)$ the monopole operator, 
	\begin{equation}\label{eq:Monpl}
	m(\tau)=e^{i\Omega\tau}\ket{e}\bra{g}+e^{-i\Omega\tau}\ket{g}\bra{e}=\begin{pmatrix}
	0&e^{+i\Omega\tau}\\
	e^{-i\Omega\tau}&0\\
	\end{pmatrix}\:,
	\end{equation}
takes the ground state of the detector to the excited state and vice versa, and can be interpreted physically as the click in response to the presence of the field. This Hamiltonian, eq.~(\ref{eq:IntH}), is essentially written in the interaction picture~\cite{Sachs2017,Arias2018} which later allows us to use expectation values of the free fields.
 
As stated earlier, the operator $\mathcal{O}_\text{F}$ describes how the quantum field couples to the detector. The usual coupling to a scalar field $\phi$ is simply linear
	\begin{equation}\label{eq:SCoup}
	\mathcal{O}_\phi[x(\tau)]=\phi[x(\tau)]\;.
	\end{equation} 
In principle the detector has finite size and so this coupling would be smeared over that volume; however we will focus on point-like detectors for ease of calculation.
 
When considering quadratic scalar and fermionic detector couplings there are persistent divergences that cannot be regulated either by the switching function $\chi_\mathcal{T}(\tau)$  or by smearing the detector~\cite{Takagi1986,Hummer2016,Louko2016Fermions}. However these can be regulated by the standard field theoretic technique of normal ordering~\cite{Hummer2016}, in which $:\hat{A}:\;=\hat{A}-\bra{0}\hat{A}\ket{0}$ for an operator $\hat{A}$.  The properly regulated quadratic couplings we consider are
	\begin{subequations}
	 \begin{alignat}{2}
	\mathcal{O}_{\phi^2}[x(\tau)]&=\;:\phi^2[x(\tau)]:\label{eq:QCoup} \\
	\mathcal{O}_\Psi[x(\tau)]&=\;:\overline{\Psi}[x(\tau)]\Psi[x(\tau)]:\;  \label{eq:FCoup}
	\end{alignat}
	\end{subequations}
where $\Psi$ is a fermionic spinor field and $\phi$ the same scalar field as in the linear case (see Appendix~\ref{App:QFT Con} for details).
Note that the linear coupling eq.~(\ref{eq:SCoup}) is trivially normal ordered. 

One final point must be made before moving on. Since each of the operators $\mathcal{O}_F$ has a different dimensionality so does each coupling constant $\lambda_F$. In fact for dimensionless switching functions one can show
	\begin{equation}\label{eq:CoupDim}
	[\lambda_F]\equiv-\Delta_F=\begin{cases}
							(4-d)/2 &\text{for } F=\phi\\
							3-d	& \text{for } F=\phi^2\\
							2-d & \text{for } F=\Psi\\
						   \end{cases}
	\end{equation}
in units where $[x]=-1$. Note that in the real world case of $d=4$ only the linearly coupled scalar is dimensionless. However one can write a dimensionless constant $\tilde{\lambda}_F=\lambda_F/\xi^{\Delta_F}$ where $\xi$ is a time scale of any perturbative approximation to the evolution of the detector.  This dimensionality will have significant consequences when we try to construct the thermodynamic cycle.

\subsection{Evolution of the Detector}

We take the initial state of the detector and the field to be the product state $\varrho_0=\rho_0\otimes\ket{0}\bra{0}$ where $\ket{0}$ is the vacuum of the scalar field in an inertial frame and $\rho_0$ is the initial density matrix describing the two level system
	\begin{equation}\label{eq:Rho0}
	\rho_0=p\ket{e}\bra{e}+(1-p)\ket{g}\bra{g}=\begin{pmatrix}
	p&0\\
	0&1-p
	\end{pmatrix}\;,
	\end{equation}
with $p$ the probability to be in the excited state. After the evolution we will trace out the scalar field as we are only interested in the response of the two level system. This is in the same spirit as allowing the density matrix of the two level system to be in contact with, and then isolating it from, a thermal bath. Now the evolved density matrix after interaction over the timescale $\mathcal{T}$ is given by
	\begin{equation}\label{eq:rho(t)}
	\varrho_\mathcal{T}=U_\mathcal{T}\;\varrho_0\;U^\dagger_\mathcal{T}\;,
	\end{equation}
where $U_\mathcal{T}$ is the time evolution operator of the interaction Hamiltonian governed by the equation
	\begin{equation}\label{eq:Liouv2}
	i\partial_\tau\;U_\mathcal{T}=\mathcal{H}_\mathrm{Int}(\tau)\;U_\mathcal{T}\;.
	\end{equation}
This is solved with a time ordered exponential given pertubatively by the Dyson series
	\begin{equation}\label{eq:SerSol}
	U_\mathcal{T}=\mathds{1}-i\int^\infty_{-\infty} \dd{\tau}\mathcal{H}^I_\mathrm{Int}(\tau)-\frac{1}{2}\int^\infty_{-\infty} \dd{\tau}\int^\infty_{-\infty}\dd{\tau'}T\left\{\mathcal{H}^I_\mathrm{Int}(\tau)\mathcal{H}^I_\mathrm{Int}(\tau')\right\}+O(\lambda_F^3) 
	\end{equation} 
with $T$ denoting time ordering $T\{A(t)B(t')\}=\Theta(t-t')A(t)B(t')+\Theta(t'-t)B(t')A(t)$. We can substitute eq.~(\ref{eq:SerSol}) into our expression for the evolved density matrix eq.~(\ref{eq:rho(t)}) and working consistently to second order we obtain
	\begin{align}\label{eq:SerSol3}
	\varrho_\mathcal{T}=\varrho_0&-i\int^\infty_{-\infty}\dd{\tau}\big[\mathcal{H}^I_\mathrm{Int}(\tau),\varrho_I\big]\nonumber\\
	&+\frac{1}{2}\int^\infty_{-\infty}\dd{\tau}\int^\infty_{-\infty}\dd{\tau'}\Big[2\mathcal{H}_\mathrm{Int}(\tau')\varrho_0\mathcal{H}_\mathrm{Int}(\tau)\nonumber\\
	&-T\big\{\mathcal{H}_\mathrm{Int}(\tau)\mathcal{H}_\mathrm{Int}(\tau')\varrho_0-\varrho_0\mathcal{H}_\mathrm{Int}(\tau)\mathcal{H}_\mathrm{Int}(\tau')\big\}\Big]+O(\lambda_F^3)\;.
	\end{align}
As only the degrees of freedom of the detector are of interest,  we take a partial trace over the field. That is to say the evolved state of the detector is given by
	\begin{equation}
	\rho_\mathcal{T}=\mathrm{Tr_{Field}}[\varrho_\mathcal{T}]=\;\mathclap{\displaystyle\int}\mathclap{\textstyle\sum}\;\;\bra{k}\varrho_\mathcal{T}\ket{k}\;,
	\end{equation}
where $k$ labels the (discrete or continuous) states of the field. Clearly $\mathrm{Tr_{Field}}[\varrho_0]=\rho_0$. Further the first order term vanishes since the vacuum expectation value of the field operator vanishes. That is $\bra{0}\mathcal{O}_\text{F}[x(\tau)]\ket{0}=0$ for all the couplings by virtue of the normal ordering. 

Thus the first non trivial contribution comes at second order. We find the density matrix of the system after time $\mathcal{T}$ of contact with the vacuum becomes 
	\begin{equation}\label{eq:Rhot}
	\rho_\mathcal{T}=\begin{pmatrix}
	p+\delta p_\mathcal{T}&0\\
	0&1-p-\delta p_\mathcal{T}\\
	\end{pmatrix}+O(\lambda_F^4)\;,
	\end{equation}
where the vacuum fluctuations give rise to a change in population of the detector 
	\begin{align}\label{eq:PopChange}
	\delta p_\mathcal{T}&=\lambda_F^2\int^\infty_{-\infty}\dd{\tau}\int^\infty_{-\infty}\dd{\tau'}\chi_\mathcal{T}(\tau)\chi_\mathcal{T}(\tau') ((1-p)e^{-i\Omega (\tau-\tau') }-pe^{i\Omega (\tau-\tau') })\;\mathcal{W}_F( \tau,\tau' )\nonumber\\
	&=\lambda_F^2\left[(1-p)\mathcal{F}_F(\Omega,\mathcal{T})-p\mathcal{F}_F(-\Omega,\mathcal{T})\right]\;.
	\end{align}
Here 
	\begin{equation}\label{eq:Resp}
		\mathcal{F}_F(\Omega,\mathcal{T})=\int_{-\infty}^{+\infty} \dd{\tau} \int_{-\infty}^{+\infty} \dd{\tau'}\chi_\mathcal{T}(\tau)\chi_\mathcal{T}(\tau')
		\mathcal{W}_F(\tau,\tau')e^{-i\Omega (\tau-\tau') }
	\end{equation}
is the response function of the detector~\cite{Birrell1984quantum,Unruh1976,DeWitt79} which is proportional to the probability for the detector to transition from ground state to excited state. We have also introduced the notation $\mathcal{W}_F(\tau,\tau')$
for the vacuum expectation value
	\begin{equation}\label{eq:VacCor}
	\mathcal{W}_F(\tau,\tau')=\bra{0}\mathcal{O}_\text{F}[x(\tau)]\mathcal{O}_\text{F}[x(\tau')]\ket{0}\;.
	\end{equation}

In appendix \ref{App:VCF} we evaluate this vacuum correlator, eq.~(\ref{eq:VacCor}), for the different couplings. For the linear coupling to the massless scalar field (eq.~(\ref{eq:SCoup})),  this is simply the Wightman function
	\begin{equation}\label{eq:SExp}
	\mathcal{W}_\phi(\tau,\tau')=\bra{0}\phi[x(\tau)] \phi[x(\tau')]\ket{0}\;.
	\end{equation}
For the quadratic coupling, eq.~(\ref{eq:QCoup}), we find as in refs.~\cite{Hummer2016,Sachs2017}
	\begin{equation}\label{eq:QExp}
	\mathcal{W}_{\phi^2}(\tau,\tau')=\bra{0}:\phi^2[x(\tau)]:\;:\phi^2[x(\tau')]:\ket{0}=2\left[\mathcal{W}_\phi(\tau,\tau')\right]^2 
	\end{equation}
a result that essentially   follows from Wick's theorem~\cite{Peskin1995QFT}; for the explicit proof see appendix~\ref{App:VCF}. 
Finally for the massless fermionic coupling, eq.~(\ref{eq:FCoup}),  we find (following~\cite{Louko2016Fermions})
	\begin{equation}\label{eq:FExp}
	\mathcal{W}_\Psi(\tau,\tau')=\bra{0}:\overline{\Psi}\Psi[x(\tau)]:\;:\overline{\Psi}\Psi[x(\tau')]:\ket{0}=\frac{N_d\Gamma(d/2)^2}{\Gamma(d-1)}\mathcal{W}^{2d}_\phi(\tau,\tau')
	\end{equation}
where $N_d=2^{d/2}$  in even ($2^{(d-1)/2}$ in odd) dimensions and $\mathcal{W}^{2d}_\phi(\tau,\tau')$ is the Wightman function of the massless scalar field in $2d$ dimensions (see appendix~\ref{App:VCF}). 

Surprisingly, all of the couplings considered depend only on this scalar Wightman function on which the population change in eq.~(\ref{eq:PopChange}) depends. Moreover, this population change will determine the thermodynamic properties of the Unruh quantum Otto cycle. In Appendix~\ref{App:EvalResp} we analytically calculate this for each of the couplings considered using a Lorentzian switching function.

Having described the dynamics of the different UdW models, we now briefly review the Unruh effect in the context of the response of a uniformly accelerated detector.

\subsection{Accelerated observers and the Unruh effect}\label{sec:AccObs}
First, to motivate the use of the Unruh effect as a thermal bath for a heat engine/cycle let us discuss thermality and see in what sense the Unruh--DeWitt detector sees a thermal vacuum. Typically in quantum mechanics one says that thermal/equilibrium states are represented by Gibbs states, i.e. states of maximum entropy~\cite{Strocchi2005,Garay2016}. However an alternative definition is provided by the Kubo--Martin--Schwinger (KMS) condition~\cite{Kubo1957,Schwinger1959}.
Consider the pullback of a quantum field $\varphi[x(\tau)]$ along a worldline $x(\tau)$ generated by the timelike vector $\partial_\tau$. Then the KMS condition on the field configuration $\rho$ implies that Wightman function of the field obeys~\cite{Fewster2016}
	\begin{equation}\label{eq:KMS}
	\mathcal{W}(\Delta\tau-i\beta)=(-1)^{2S}\mathcal{W}(-\Delta\tau)\;.
	\end{equation}
where $S=1\;(1/2)$ for bosons (fermions)~\cite{Birrell1984quantum}.   

All Gibbs states are KMS states~\cite{Garay2016} and KMS states are passive (i.e. cannot do work). So the KMS condition reproduces our standard notion of thermality. However not all KMS states are Gibbs states~\cite{Garay2016}. To be in equilibrium generally has extra requirements such as stability. Moreover it can be shown eq.~(\ref{eq:KMS}) implies that the Fourier transform of the Wightman function of a KMS state is proportional to the Bose/Fermi distribution with inverse temperature $\beta$~\cite{Fewster2016,Garay2016}
	\begin{equation}
	\hat{\mathcal{W}}(\Omega)\propto\frac{1}{1-(-1)^{2S}e^{\beta\Omega}}\;.
	\end{equation}

Now let us consider a scalar field along the worldine of a uniformly accelerated observer. In Appendix~\ref{App:VCF} we show that the Wightman function of a massless scalar field in the $d$-dimensional Minkowski vacuum takes the form
	\begin{equation}\label{eq:SWight}
	\mathcal{W}_\phi(\tau,\tau')=\frac{\Gamma(d/2-1)}{4\pi^{d/2}}[z(\tau,\tau')]^{2-d}\;,
	\end{equation}
where $\Gamma(z)\equiv\int_{0}^{\infty}t^{z-1}e^{-t}\dd{t}$ is the gamma function and
	\begin{align}
	z(\tau,\tau')&=\epsilon+i\;\text{sgn}(x^0-x'^0)\Delta(\tau,\tau')\;,\\
	\Delta(\tau,\tau')&=|x(\tau)-x(\tau')|\;.
	\end{align}
The Wightman function is a distribution and is understood in the usual limit $\epsilon\rightarrow 0^+$.
Now the world line of a uniformly accelerated observer is described by the hyperbolic trajectory
	\begin{equation}\label{eq:Wline}
	t(\tau)=\frac{1}{\alpha}\sinh(\alpha\tau)\;,\;x(\tau)=\frac{1}{\alpha}\cosh(\alpha\tau)
	\end{equation}
yielding for the Wightman function  
	\begin{equation}\label{eq:SWight2}
	\mathcal{W}_\phi(\tau,\tau')=\frac{\Gamma(\frac{d}{2}-1)}{4\pi^{d/2}}\left(\frac{\alpha}{2i\sinh[(\alpha \Delta\tau-i\epsilon)/2]}\right)^{d-2}\;,
	\end{equation}
where we have absorbed a positive function of $\Delta\tau$ into the infinitesimal $\epsilon$.
From the periodicity of the $\sinh(z)$ function it is clear that, for the pullback of the scalar field along an accelerated worldline, the Wightmann function satisfies the KMS condition eq.~(\ref{eq:KMS})
	\begin{eqnarray}\label{eq:KMSAcc}
	\mathcal{W}_\phi( \Delta\tau+i\beta_U)=(-1)^{d}\;\mathcal{W}_\phi(- \Delta\tau)\;;\;\beta_U=2\pi/\alpha\;,
	\end{eqnarray}
for a thermal state with temperature $T_U=\beta_U^{-1}$. Notice that this corresponds to a Bose distribution in even dimensions and a Fermi one in odd dimensions~\cite{Takagi1986}. 

The question remains: what will an accelerated observer see due to the fluctuations of the quantum field? This is answered with the response function of the Unruh--DeWitt detector. In the long time limit the response function  (\ref{eq:Resp}) is proportional to the Fourier transform of the Wightman function~(\ref{eq:SWight2}) ~\cite{Fewster2016,Garay2016}
	\begin{equation}\label{eq:LongT}
	\lim_{\mathcal{T}\rightarrow\infty}\mathcal{F}(\Omega,\mathcal{T})/\mathcal{T}\propto\hat{\mathcal{W}}_\phi(\Omega) 
	\end{equation}
and so a uniformly accelerated observer sees  a thermal state. This is the manifestation of the Unruh effect  understood in a precise way using the KMS condition. 

For the quadratically coupled scalar and fermion case the limit will change to be the Fourier transform of the respective vacuum correlators~(\ref{eq:QExp}, \ref{eq:FExp}). As these depend only on the Wightman function, they are clearly also KMS states. Somewhat surprisingly, the response to fermions is always a Bose distribution~\cite{Louko2016Fermions} since it depends on the scalar Wightman function in $2d$ and so $\mathcal{W}_\Psi(\Delta\tau+i2\pi/\alpha)=\mathcal{W}_\Psi(-\Delta\tau)$ (see eq.~(\ref{eq:KMSAcc})). We take this as motivation for the next section, in which accelerating the UdW detector, over a finite time, plays the role of contact with an approximately thermal reservoir.

\section{The quantum Otto cycle}\label{sec:UQOC}

In this section we recapitulate the basic features of the Unruh quantum Otto cycle.

Recall first a few definitions from quantum thermodynamics. While we shall be only considering a two level system with a state described by a density matrix $\rho(t)$ and associated time dependent Hamiltonian $\mathcal{H}(t)$, the following statements are quite general (for countably infinite dimensional systems). The average energy of the system $\langle E(t)\rangle=\text{Tr}[\rho(t)\mathcal{H}(t)]$ obeys the quantum analogue of the first law of thermodynamics~\cite{Balian2006}
	\begin{equation}\label{eq:1stlaw}
	\partial_t\langle E(t)\rangle=\text{Tr}[\partial_t\rho(t)\mathcal{H}(t)]+\text{Tr}[\rho(t)\partial_t\mathcal{H}(t)]\;,
	\end{equation}
where the first term on the right hand side encodes the change in the internal states and populations of the system and the second can be interpreted as external shifting of the energy levels of the system. Thus it is natural to identify 
	\begin{equation}\label{eq:Heat}
	\langle Q\rangle=\int_\mathcal{T}\dd{t}\text{Tr}\left[\frac{\partial\rho(t)}{\partial t}\mathcal{H}(t)\right]
	\end{equation}
with the average heat transfer to the system over time $\mathcal{T}$. We can also recognise
	\begin{equation}\label{eq:Work0}
	\langle W\rangle=\int_\mathcal{T}\dd{t}\text{Tr}\left[\rho(t)\frac{\partial\mathcal{H}(t)}{\partial t}\right]
	\end{equation}
as the work done on the system over the period of interaction. 

Finally, the von Neumann entropy is $S=\text{Tr}[\rho\log\rho]$ and the equilibrium Gibbs/Boltzmann distribution is $\rho=\exp(-\beta\mathcal{H})/Z$ where $Z$ is the partition function. It then follows from eqs.~(\ref{eq:Heat}, \ref{eq:Work0}) that $T\delta S=\Tr[\delta \rho \mathcal{H}]=\delta Q$,  compatible with our usual expectations of the classical laws of thermodynamics.

\subsection{Otto cycle}

The classical Otto cycle  forms the basis of combustion engines. It is a thermodynamic process in which a system undergoes adiabatic (no heat transfer, $\delta Q=0$) compressions and expansions and two steps of pressure changes, while this system is held at constant volume. 

A quantum analogue of this cycle was first proposed in ref.~\cite{Kieu2004} for a two level (qubit) system. In this model the classical steps of adiabatic expansion and contraction of a piston are implemented by expanding and contracting the energy gap of the two level system. We could imagine this being implemented by some kind of weak electric or magnetic field which tunes the energy gap, without disturbing the state of the system. However the details of the process will not be important for our considerations.

Concretely suppose we have a system with ground state $\ket{g}$, with zero energy, and an excited state $\ket{e}$, with energy $\Omega_1$. The Hamiltonian for such a system is $H=\Omega_1\ket{e}\bra{e}$. As in section~\ref{sec:UdWmod} the Otto cycle begins with the system prepared in the state $\rho_0=p\ket{e}\bra{e}+(1-p)\ket{g}\bra{g}$. Subsequently,

\begin{enumerate}

\item The system undergoes an adiabatic expansion of the energy gap $\Omega_1$ to $\Omega_2$. No heat is exchanged with the environment but some work is required.

\item The qubit $\rho_0$ is placed in contact with a thermal bath at temperature $T_H$. After interacting for some time $\mathcal{T}_2$ heat is exchanged and the system will be in the state $\rho=(p+\delta p_H)\ket{e}\bra{e}+(1-p-\delta p_H)\ket{g}\bra{g}$. There is no work done in this step (analogous to a classical constant volume step).

\item The energy level of the qubit is contracted adiabatically from $\Omega_2$ to $\Omega_1$. As in step one, no heat is exchanged. This is the power stroke of the cycle where the system performs work.

\item In the final stage the system brought into contact with a thermal bath with temperature $T_C<T_H$. This is the cold reservoir. After time $\mathcal{T}_1$ the final state of the qubit is $\rho_f=(p+\delta p_H+\delta p_C)\ket{e}\bra{e}+(1-p-\delta p_H-\delta p_C)\ket{g}\bra{g}$. Again no work is done in this step. In order for the final state to equal the initial, such that the cycle is closed, we must have $\delta p_H+\delta p_C=0$.
\end{enumerate}

In contrast to the classical case, it is possible that heat flows from the hot reservoir to the cold i.e. $\delta p_H<0$~\cite{Kieu2004,Kieu2006}. This is because the processes are inherently probabilistic and is due to quantum, not thermal, fluctuations. Consistency with the second law of thermodynamics requires that over a large number of cycles $\langle\delta p_H\rangle>0$.

\subsection{Unruh Otto Engine}

Having seen how a qubit can be put through the quantum version of an Otto engine and that UdW detectors will see a thermal vacuum in the KMS sense, we now turn to  exploitation of the thermal vacuum for a similar Otto cycle. To achieve this we will use the UdW detector model to describe the coupling of our qubit to some background quantum field as discussed in Section~\ref{sec:UdWmod}. The kinematic and thermodynamic steps are similar to that of the linearly coupled detector~\cite{Arias2018}, but we shall review them here for completeness.
 
In steps 2 and 4 of the Otto cycle the cold and hot temperatures will correspond to different accelerations $\alpha_H$ and $\alpha_C$. Later we will see that there are precise conditions on the hot and cold accelerations such that positive work is extracted. For now we will say that it is to be expected that we must have $\alpha_H>\alpha_C$ for the hot vacuum to have higher temperature than the cold, i.e. $T_H>T_C$, and thus transfer work from the vacuum to the system.

\begin{figure}[h!]
		\begin{center}
		\includegraphics[scale=0.75]{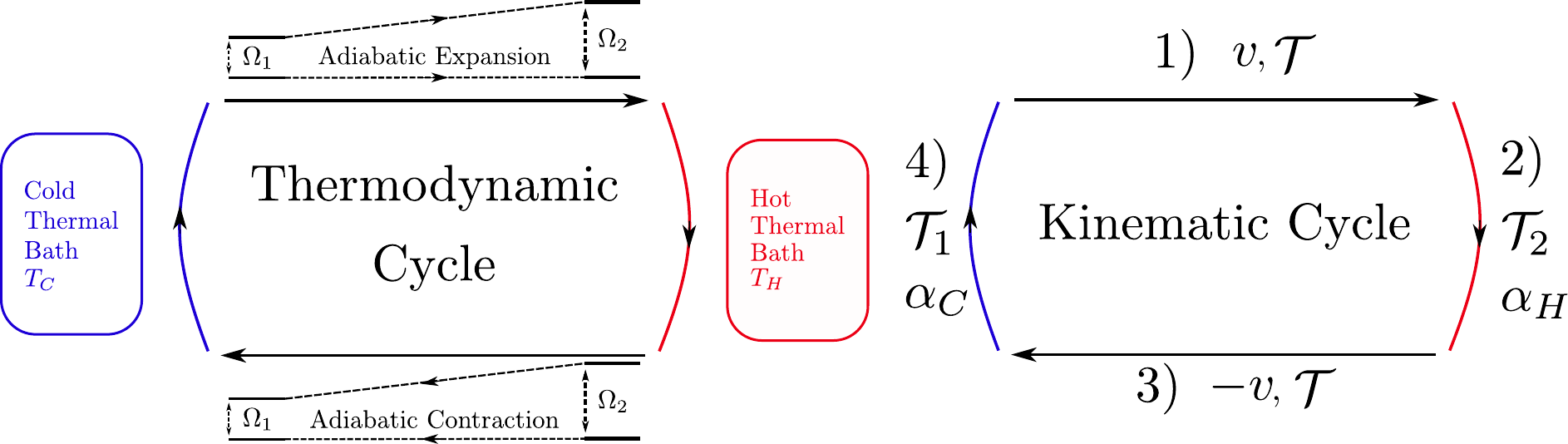}
	
		\caption[Pole Structure]{\label{fig:Cycle}\small The kinematic and thermodynamic cycles of the Unruh quantum Otto engine adapted from ref.~\cite{Arias2018}. Acceleration plays the role of temperature and contact with the quantum vacuum is analogous to a thermal bath.}
		\end{center}
	\end{figure}

In steps 1 and 3 we will assume that the qubit is travelling at constant velocity and can be isolated from the quantum field vacuum in a similar way that one can isolate any other thermal system. Moreover we require that the kinematic cycle of the qubit is closed. This leads to the following steps in the cycle depicted schematically in Figure~\ref{fig:Cycle}

\begin{enumerate}
	\item Adiabatic expansion: The qubit is travelling at constant velocity $v$ for time $\mathcal{T}$, during which the energy gap expands from $\Omega_1$ to $\Omega_2$.

	\item Hot contact: Constant acceleration $\alpha_H$ of the qubit for time $\mathcal{T}_2$ where the vacuum acts as hot thermal reservoir. The velocity of the qubit accelerates from $v$ to $-v$.

	\item Adiabatic contraction: The qubit is travelling at velocity $-v$ for	 time $\mathcal{T}$ during which the energy gap contracts from $\Omega_2$ to $\Omega_1$.

	\item Cold contact: Constant acceleration $\alpha_C$ from velocity $v$ to $-v$ for time  $\mathcal{T}_1$ where the vacuum acts as a cold reservoir and the qubit returns to its initial state.
\end{enumerate}

The times $\mathcal{T}_1$ and $\mathcal{T}_2$ for which the system accelerates and is in contact with the vacuum are fixed by the requirement that the velocity change from $v$ to $-v$ and vice versa so that it returns to its original state and the cycle is kinematically complete. A constantly accelerated observer has a hyperbolic worldline $x(\tau)$, see eq.~(\ref{eq:Wline}), and differentiating these expressions they must have speed
	\begin{equation}\label{eq:vel}
	v=\tanh(\alpha\tau)\;.
	\end{equation}
Thus the time taken to velocity $v$ from $\tau=0$ is simply given by inverting eq.~(\ref{eq:vel}): $\tau=\text{arctanh}(v)/\alpha$. Hence in steps 2 and 4, with accelerations $\alpha_H$ and $\alpha_C$ respectively, the qubit must accelerate for the following times
	\begin{align}\label{eq:Times}
	\mathcal{T}_2=2\; \text{arctanh}(v)/\alpha_H\;,\\
	\mathcal{T}_1=2\; \text{arctanh}(v)/\alpha_C\;.
	\end{align}
Having described the kinematics of the qubit's cycle we now perform a preliminary examination of the thermodynamics of each step in the cycle.

\subsubsection{Adiabatic expansion}

In this step the energy gap changes from $\Omega_1$ to $\Omega_2>\Omega_1$ over a time $\mathcal{T}$ while the state of the system $\rho_0=p\ket{e}\bra{e}+(1-p)\ket{g}\bra{g}$ is held constant. Thus we have a time dependent Hamiltonian
	\begin{equation}
	\mathcal{H}(t)=\Omega(t)\ket{e}\bra{e}\;.
	\end{equation}
From our expression for the average heat exchanged, eq.~(\ref{eq:Heat}), it is clear that this step is adiabatic, i.e.
	\begin{equation}
	\langle Q_1\rangle=\int_\mathcal{T}\dd{t}\text{Tr}\left[\frac{\partial\rho_0}{\partial t}\mathcal{H}(t)\right]=0\;.
	\end{equation}
However there is positive work done on the system
\be
	\langle W_1\rangle =\int_\mathcal{T}\dd{t}\text{Tr}\left[\rho_0\frac{\partial \mathcal{H}(t)}{\partial t}\right]
 =\int_\mathcal{T}\dd{t}\dot{\Omega}\;\text{Tr}[\rho_0\ket{e}\bra{e}] 
 =(\Omega_2-\Omega_1)p\;.
\ee 
%

\subsubsection{Contact with the hot vacuum}

In this step the Hamiltonian of the qubit is held constant $\mathcal{H}=\Omega_2\ket{e}\bra{e}$ while the system accelerates from $-v$ to $v$ over the interval $\mathcal{T}_2$ and interacts with a background quantum field. In Section~\ref{sec:UdWmod} we saw  at leading order the state of the qubit after interaction timescale $\mathcal{T}_2$ is of the form
	\begin{equation}
	\rho_{\mathcal{T}_2}=\rho_0+\delta p_H\;\sigma_z 
	\end{equation}
regardless of the coupling, 
where $\sigma_z=\ket{e}\bra{e}-\ket{g}\bra{g}$ and $\delta p_H=\delta p_{\mathcal{T}_2}$. Clearly there is no work done in this step, as the qubit Hamiltonian does not change:
	\begin{equation}
	\langle W_2\rangle=\int_{\mathcal{T}_2}\dd{t}\text{Tr}\left[\rho(t)\frac{\partial\mathcal{H}}{\partial t}\right]=0\;.
	\end{equation}
However the vacuum transfers heat to the system
\be \label{eq:Heat2}
	\langle Q_2\rangle_{\mathcal{T}_2} =\int_{\mathcal{T}_2}\dd{t}\text{Tr}\left[\frac{\partial\rho}{\partial t} \mathcal{H}\right] =\Omega_2\text{Tr}\left[\delta p_{\mathcal{T}_2}\sigma_z\ket{e}\bra{e}\right]
	=\Omega_2\delta p_H\;.
\ee
%

\subsubsection{Adiabatic contraction}

In this step the state travels at velocity $v$ and is held fixed $\rho=\rho_0+\delta p_H \sigma_z$ as the energy gap is tuned for $\Omega_2$ to $\Omega_1$. Much like the adiabatic expansion no heat is exchanged
	\begin{equation}
	\langle Q_3\rangle=0
	\end{equation} 
and work is done
	\begin{equation}
	\langle W_3\rangle=-(\Omega_2-\Omega_1)(p+\delta p_H)\;.
	\end{equation}
%

\subsubsection{Contact with the cold vacuum}

In the final step the Hamiltonian of the qubit is held constant $\mathcal{H}=\Omega_1\ket{e}\bra{e}$. While the system accelerates from $v$ to $-v$ over the interval $\mathcal{T}_1$ and interacts with a background quantum field. As before at leading order we have
	\begin{equation}
	\rho_{\mathcal{T}_1}=\rho_1+\delta p_C\;\sigma_z\;,
	\end{equation}
where $\delta p_C=\delta p_{\mathcal{T}_1}$, $\rho_1=p'\ket{e}\bra{e}+(1-p')\ket{g}\bra{g}$ and $p'=p+\delta p_H$. We find that no work is done
	\begin{equation}
	\langle W_4\rangle=0
	\end{equation}
and the average heat transfer is
	\begin{equation}\label{eq:Heat4}
	\langle Q_4\rangle=\Omega_1\delta p_C\;.
	\end{equation}
	\begin{figure}[htpb]
		\begin{center}
		\includegraphics[scale=0.4]{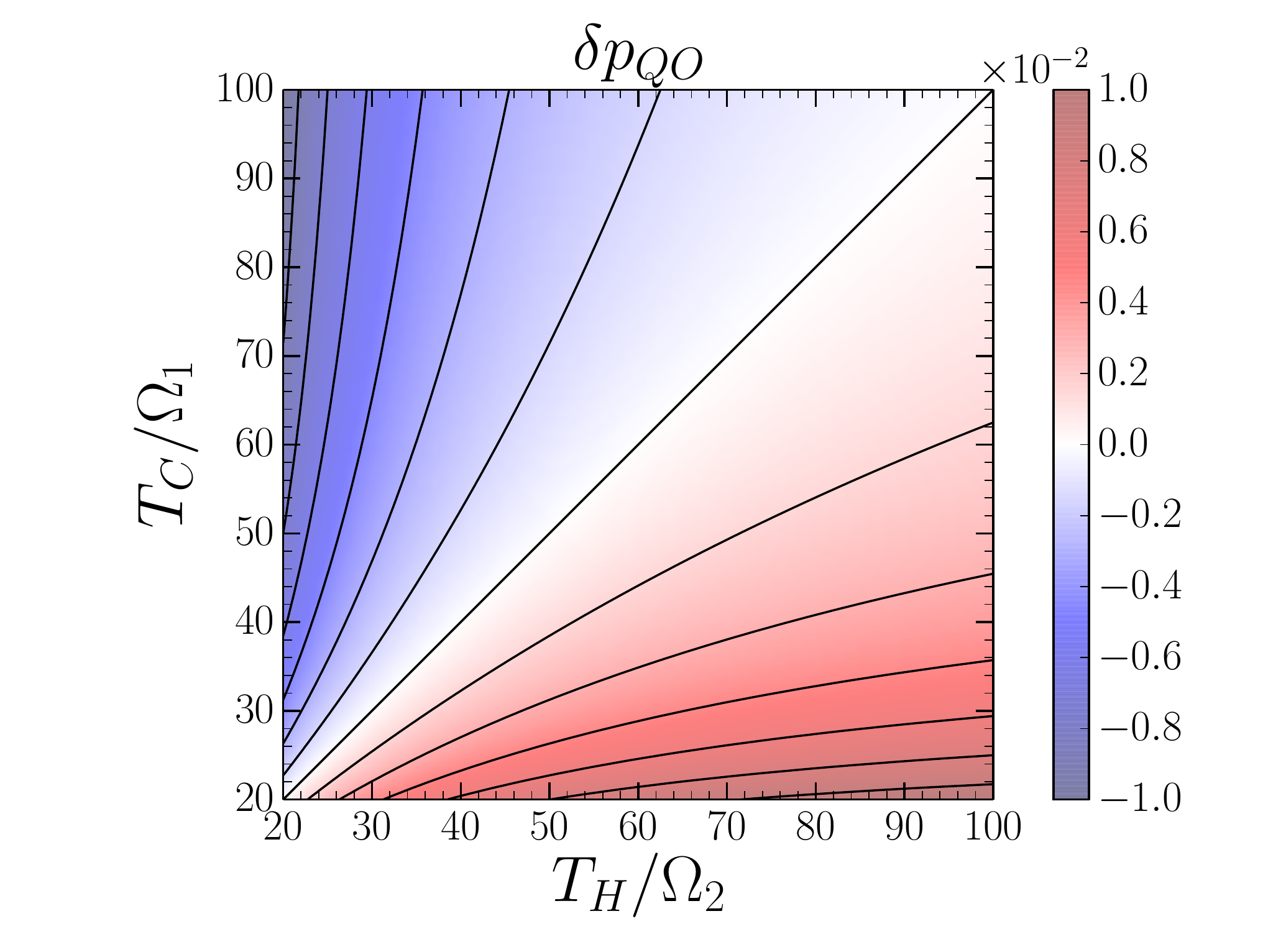}
	
		\caption[Quantum Otto Work]{\label{fig:WorkQO}\small The population change for the original quantum Otto cycle. Red regions show positive corrections and blue negative}
		\end{center}
	\end{figure}

\subsection{Completing the Cycle}

For the system to return to its original state we must have $\delta p_H+\delta p_C=0$. To complete the thermodynamic analysis the net work done by the cycle is
	\begin{equation}\label{eq:Work}
	\langle W_\text{tot}\rangle=\langle W_1\rangle+\langle W_3\rangle=-(\Omega_2-\Omega_1)\delta p_H\;,
	\end{equation}
while the heat is 
	\begin{equation}
	\langle Q_\text{tot}\rangle=\langle Q_2\rangle+\langle Q_4\rangle=(\Omega_2-\Omega_1)\delta p_H\;.
	\end{equation}
Thus, as required, we have conservation of energy:
	\begin{equation}
	\langle W_\text{tot}\rangle+\langle Q_\text{tot}\rangle=0\;.
	\end{equation}

Note that $\delta p_H$ must be positive for the external work $\langle W_{\text{ext}}\rangle=-\langle W_\text{tot}\rangle$ (i.e. the work done by the detector) to be positive. For the normal quantum Otto cycle~\cite{Kieu2004},  at the end of step 2 after thermalisation with the hot bath, the state satisfies\\ $p+\delta p_H=\Tr[\ket{e}\bra{e}\rho]=1/(1+\exp(\Omega_2/T_H))$. Imposing that the system returns to its original state means that, at the end of step 4 after thermalising with the cold bath, $p=\Tr[\ket{e}\bra{e}\rho]=1/(1+\exp(\Omega_2/T_C))$. Thus for the original quantum Otto cycle
	\begin{equation}\label{eq:Wcl}
		\delta p_{QO}=W_{QO}/(\Omega_2-\Omega_1)=\frac{1}{1+e^{\Omega_2/T_H}}-\frac{1}{1+e^{\Omega_1/T_C}}\;.
	\end{equation}
This is shown in Figure~\ref{fig:WorkQO}.

Therefore in order to get positive work we have the requirement
\begin{equation}\label{eq:PWCon}
T_H/\Omega_2>T_C/\Omega_1
\end{equation}
which is stronger than the classical counterpart $T_H>T_C$. In Section~\ref{sec:Results} we shall explore this for the Unruh quantum Otto cycle. 

\section{Thermodynamic analysis}\label{sec:Results}
Returning to the Unruh quantum Otto cycle, the time the detector interacts with the vacuum is given by $\mathcal{T}=2\text{arctanh}(v)/\alpha$  (eq.~(\ref{eq:Times})). Hence defining the reduced acceleration $a=\alpha/\Omega=1/x$ we can write the change in population of the qubit over steps 2 or 4 in the cycle (c.f. eq.~(\ref{eq:PopChange})) as
	\begin{equation}\label{eq:PopChange2}
		\delta p_F(a,p,v)=\lambda_F^2\left((1-2p)\mathcal{F}_F(1/a,2\text{arctanh}(v))-p\;\Delta\mathcal{F}_F(1/a,2\text{arctanh}(v))\right)\;.
	\end{equation}
where $\Delta \mathcal{F}_F(x,y)=-\mathcal{F}_F(x,y)+\mathcal{F}_F(-x,y)$. In Appendix \ref{App:EvalResp} we provide the analytic expressions for   the response functions for each UdW coupling.

Now in order for the perturbative approximation to be valid it must hold that $\lvert\delta p_F\rvert\ll1$. Thus for small reduced accelerations our approach may break down. To quantify this, consider that the integral expressions, eqs.~(\ref{eq:PopChange}, \ref{eq:PopChange2}), lead to the relation 
	\begin{equation}\label{eq:popdim}
	\lvert\delta p_F\rvert\sim(\lambda_F/\mathcal{T}^{\Delta_F})^2\Omega\mathcal{T}\;.
	\end{equation}
 This means that the perturbative regime is valid provided
	\begin{equation}\label{eq:PertCon}
	a/\text{arctanh}(v)\gg\left(\lambda_F/\mathcal{T}^{\Delta_F}\right)^2\;\iff\;\left(a/\text{arctanh}(v)\right)^{1-2\Delta_F}\gg\lambda_F^2\Omega^{2\Delta_F}\;.
	\end{equation}
Here we have a manifestation of the dimensionality of the coupling constant. For the linearly coupled scalar, in $d=4$, $\Delta_F=0$ and so we have an easily defined perturbation regime $a/\text{arctanh}(v)\gg\lambda_\phi^2$. However for the quadratic and fermionic case there are two time scales in the problem, $\mathcal{T}_{1/2}$, which depend on the cold and hot accelerations $\alpha_{C/H}$ respectively. Since in $d=4$, $\Delta_{\phi^2}=1$ and $\Delta_\Psi=2$ (c.f eq.~\ref{eq:CoupDim}), for fixed $v<1$, the coupling constants, $\lambda_F\Omega^{\Delta_F}$, must be much smaller in order to access high acceleration/temperature regimes. Moreover, there are also two gap sizes $\Omega_{1/2}$ that enter into the cycle. This means that in the quadratic and fermionic cases we will be required to fix the ratio of gap sizes in order to plot results. Note that by construction $\Omega_2/\Omega_1>1$.

Let us now explore the behaviour of this population change for each of our cases. Our analysis, by and large, follows ref.~\cite{Arias2018} and for the linearly coupled scalar case we essentially reproduce their results.

\subsection{Population change}

\begin{figure}[h!]
        \centering
        \captionsetup[subfigure]{labelformat=empty}
        \begin{subfigure}[h!]{0.475\textwidth}
            \centering
            \includegraphics[width=\textwidth]{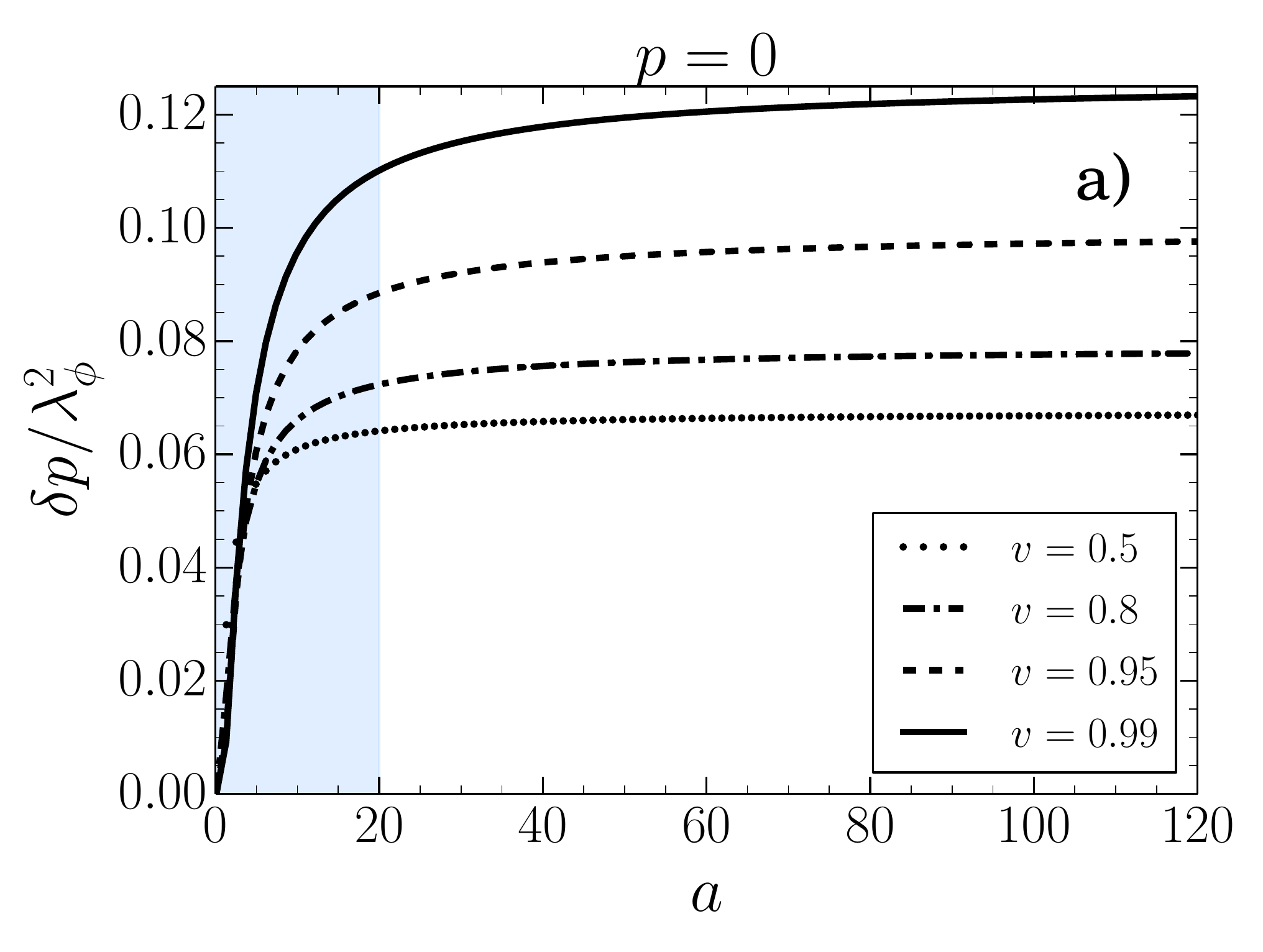}
            \caption{}
            \label{fig:p0Scalar4d}
        \end{subfigure}
        \begin{subfigure}[h!]{0.475\textwidth}  
            \centering 
            \includegraphics[width=\textwidth]{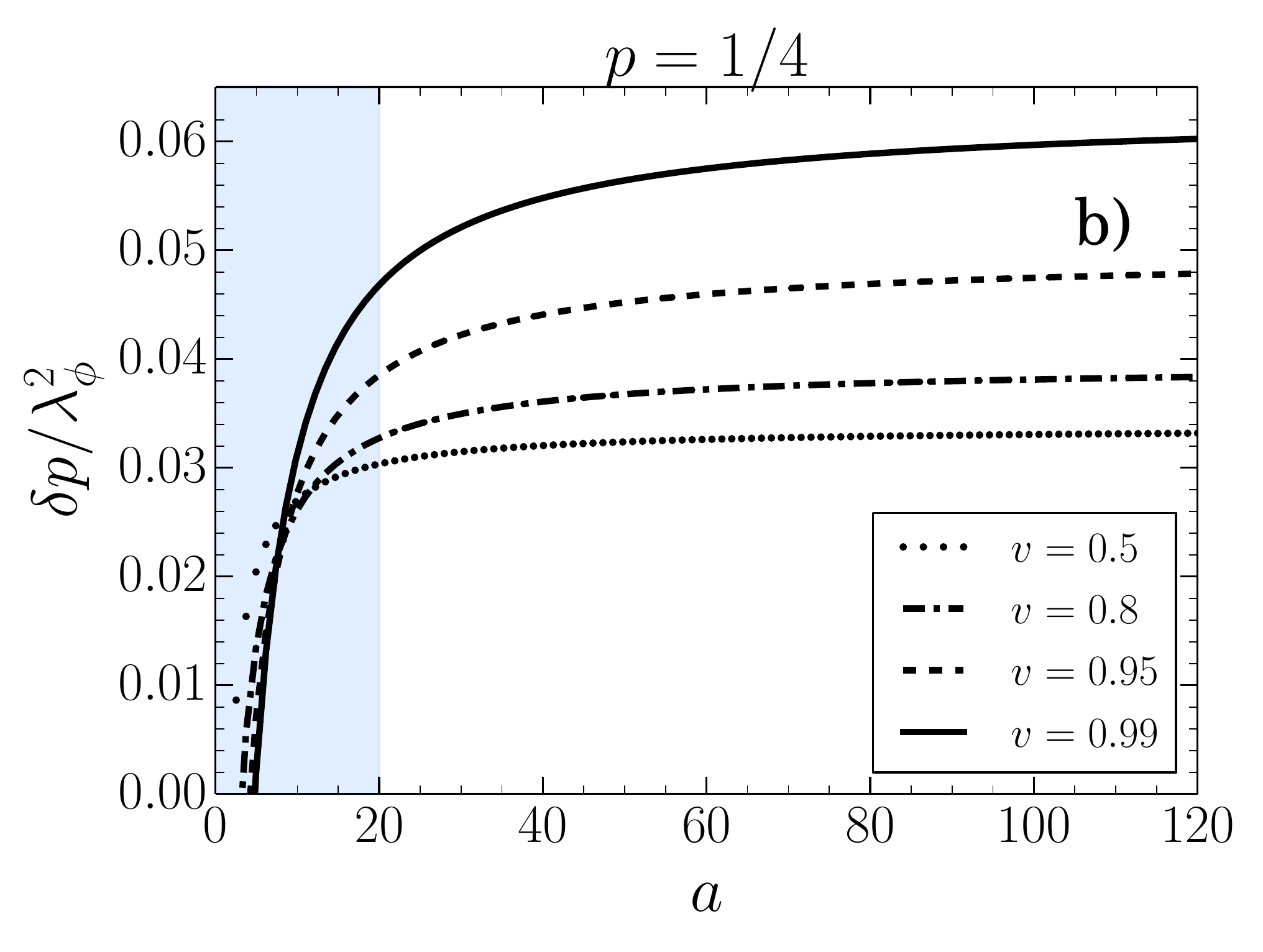}
            \caption{}
            \label{fig:p25Scalar4d}
        \end{subfigure}
        \begin{subfigure}[h!]{0.475\textwidth}   
            \centering 
            \includegraphics[width=\textwidth]{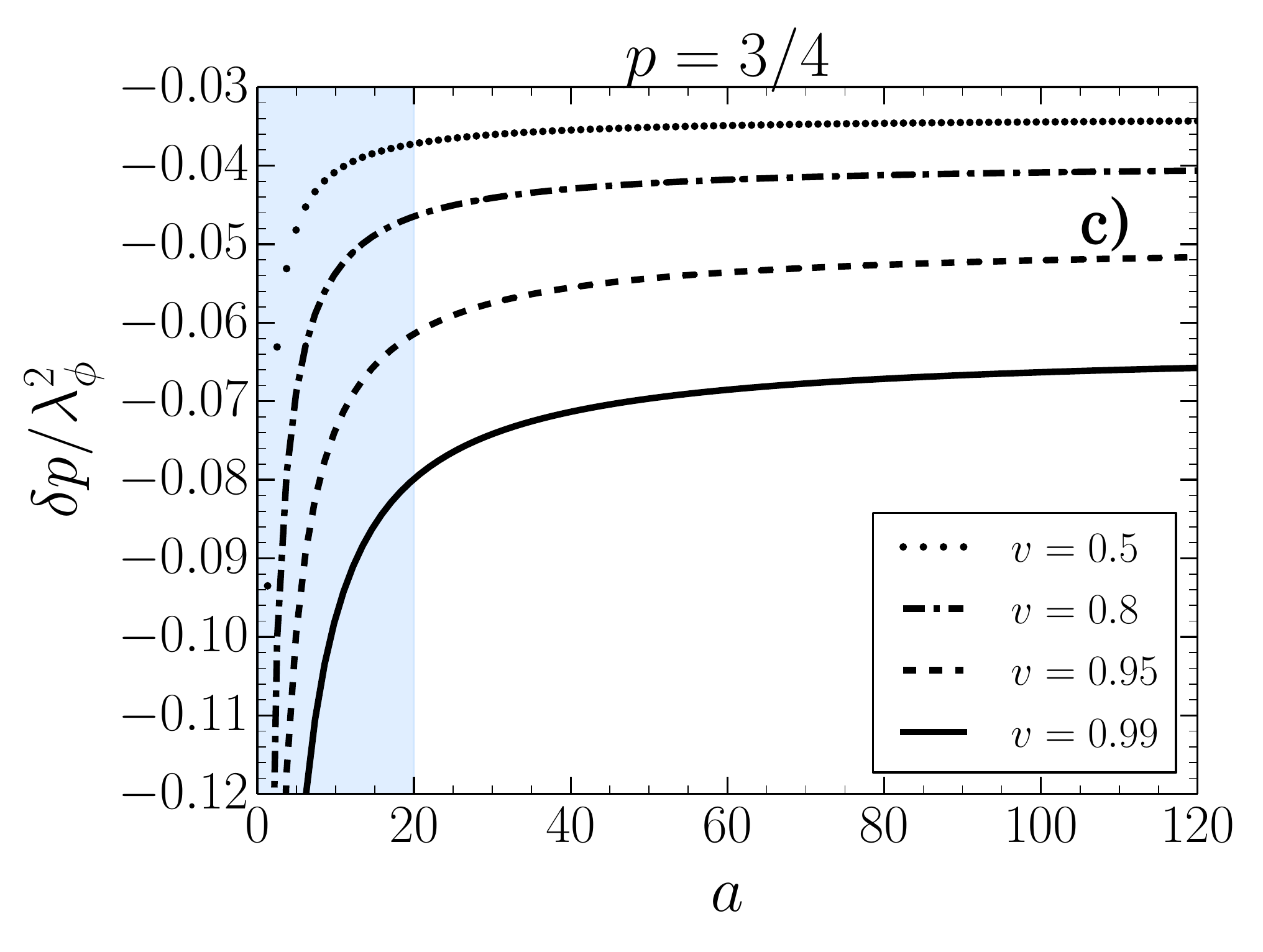}
            \caption{}
            \label{fig:p75Scalar4d}
        \end{subfigure}
        \begin{subfigure}[h!]{0.475\textwidth}   
            \centering 
            \includegraphics[width=\textwidth]{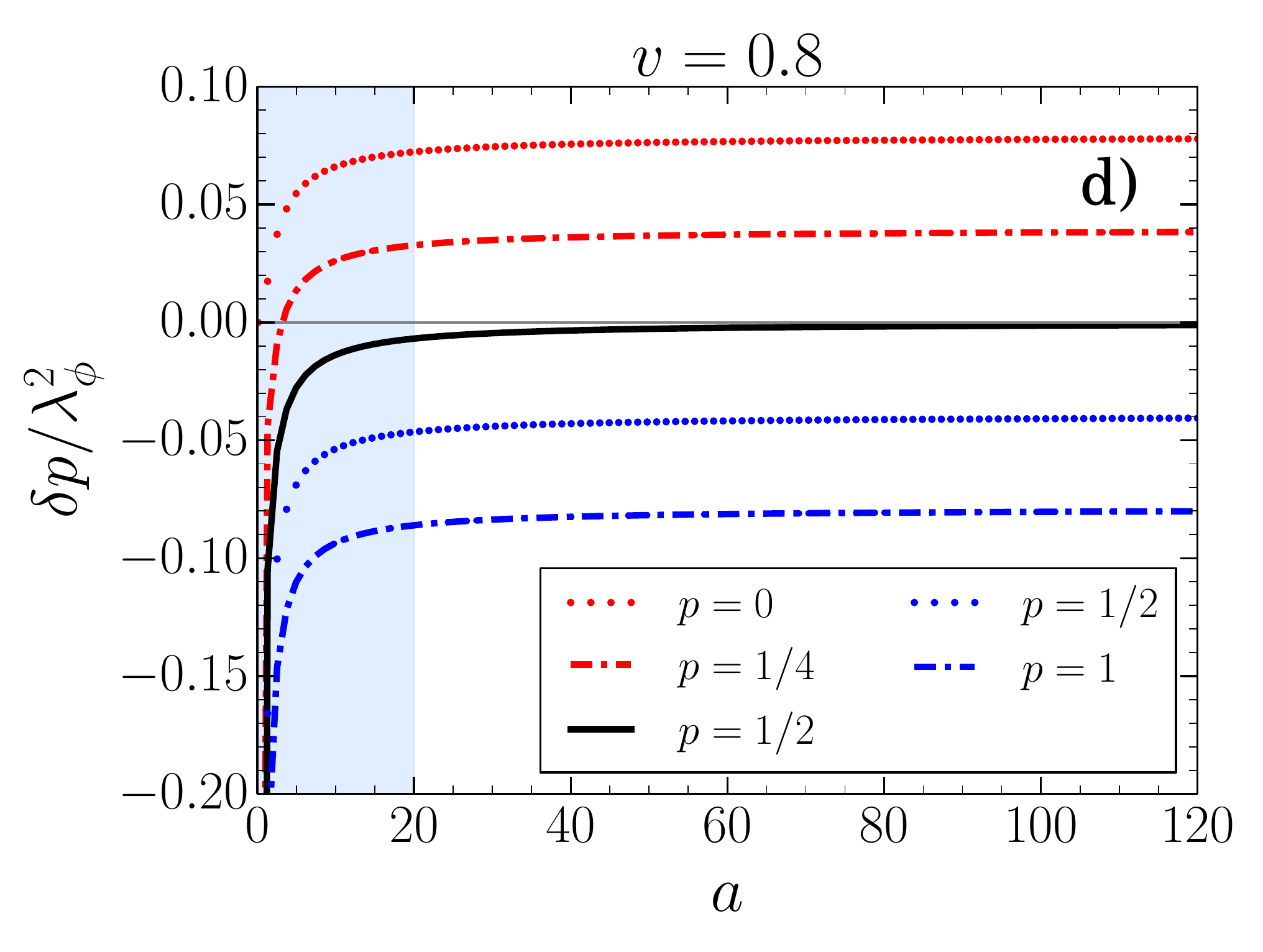}
            \caption{}
            \label{fig:v8Scalar4d}
        \end{subfigure}
        \caption[ ]
        {\small Behaviour of the change in probability for the linear scalar interaction, eq.~(\ref{eq:PopChange2}), as a function of the reduced acceleration $a=\alpha/\Omega$. Plots a), b), and c) show the effect of increasing the velocity (effectively the interaction time) on population change for a given initial excited population $p$. Plot d) shows the behaviour for a fixed velocity as the velocity changes. The shaded blue region indicates where the perturbative approximation breaks down.} 
        \label{fig:dpScalar4d}
    \end{figure}

In Figure~\ref{fig:dpScalar4d} we plot the population change eq.~(\ref{eq:PopChange2}) as a function of the reduced acceleration, $a=\alpha/\Omega$, for the linear scalar case.
We see that $\delta p_\phi$ generically increases with $a$, which corresponds to increasing temperature, and thus agrees with the intuitive expectation. Further, for all cases, the population change asymptotes to a constant determined by the velocity. The shaded blue region in these figures corresponds roughly to the region where the perturbative scheme breaks down for small $a$. In this region the $\delta p_\phi$ diverges for $p\neq0$.

Figure~\ref{fig:p0Scalar4d} shows eq.~(\ref{eq:PopChange2}) for an initially completely unexcited state (i.e. $p=0$). Generically for longer interaction time with the quantum vacuum corresponding to larger velocity $v$, we find that the correction $\delta p_\phi$ is larger. The correction is also always positive, corresponding to only excitation of the qubit/detector.

This is still true for the case of initial excited probability $p=1/4$ shown in Figure~\ref{fig:p25Scalar4d}. We can see that the change is still always positive corresponding with expectation. Since the initial state of the detector is mostly in the ground state and so interaction with the thermal vacuum causes excitation. However, one has to be a little careful with this intuition, since we are no longer dealing with eigenstates of the full Hamiltonian, and so an excitation from the ground state of the detector does not necessarily correspond to the semi-classical picture of the absorption of a particle. In this sub-plot we see clearly that the perturbation scheme breaks down: lines corresponding to different velocities cross (yielding the same population change for different interaction times, which has no physical meaning) and the divergence as $a\rightarrow0$ is clear.  This simply comes from the fact that to accelerate the qubit from $-v$ to $v$ with  vanishing acceleration takes an infinite time.

In Figure~\ref{fig:p75Scalar4d} we plot an initially mostly excited detector state with $p=3/4$. In this case interaction with the Unruh vacuum always causes a stimulated de-excitation of the detector. Finally in Figure~\ref{fig:v8Scalar4d}, we show an overlay of different initial excitation probabilities for a fixed velocity $v=0.8$. In each case it is clear that for $0\leq p<1/2$ we get positive corrections whereas we only get negative corrections for $1>p>1/2$. For the critical case of $p=1/2$ the change is always negative but approaches zero for large $a$, and an equilibrium between excitation and de-excitation is reached. Thus in order to get positive work we must have an initial state with excitation probability $0\leq p<1/2$ which, from eq.~(\ref{eq:PopChange2}), will increase linearly with $p$.

\begin{figure}[htpb]
        \centering
        \begin{subfigure}[b]{0.475\textwidth}
            \centering
            \includegraphics[width=\textwidth]{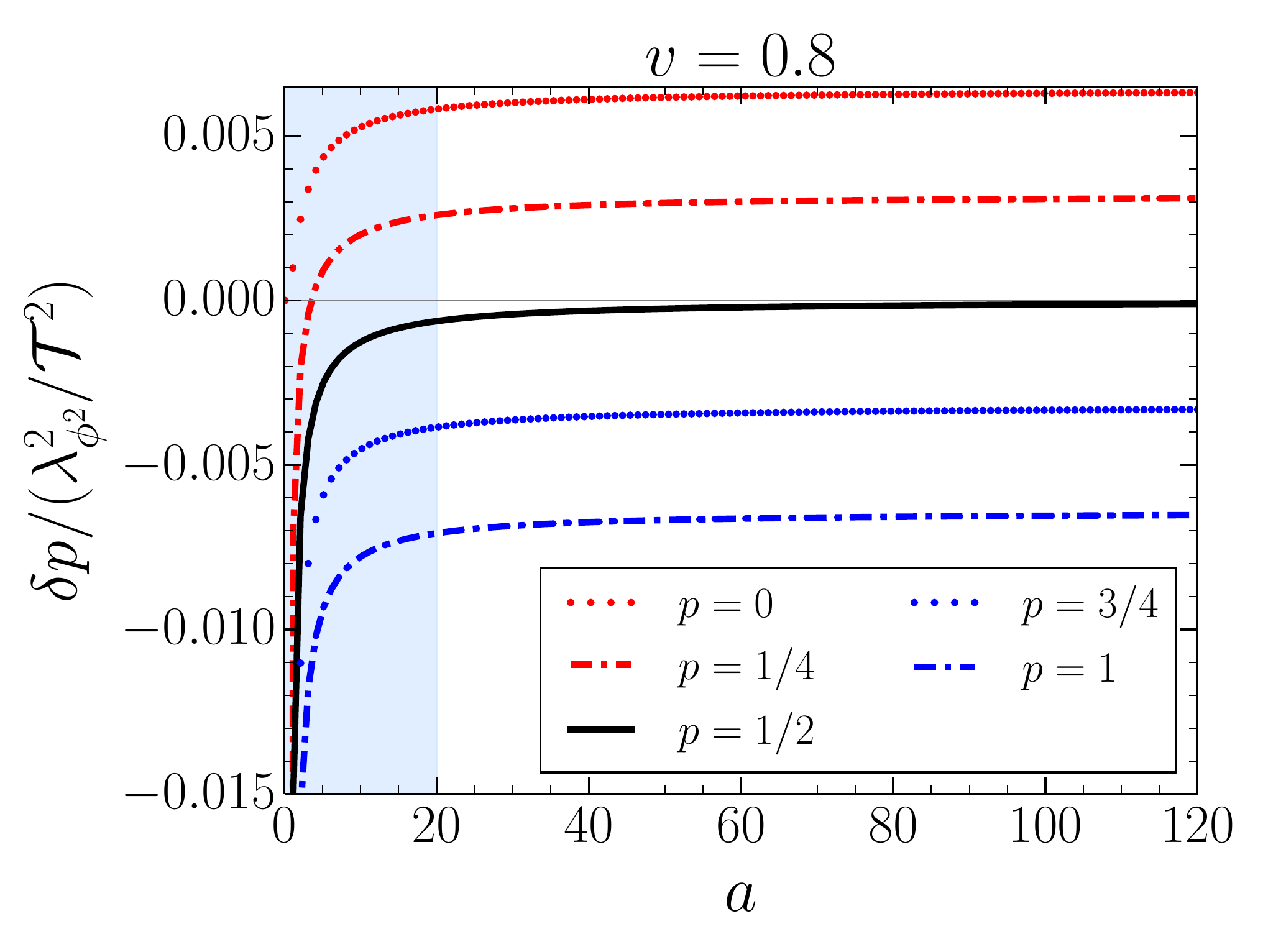}
            \caption{}
            \label{fig:v8QuadT}
        \end{subfigure}
        \begin{subfigure}[b]{0.475\textwidth}  
            \centering 
            \includegraphics[width=\textwidth]{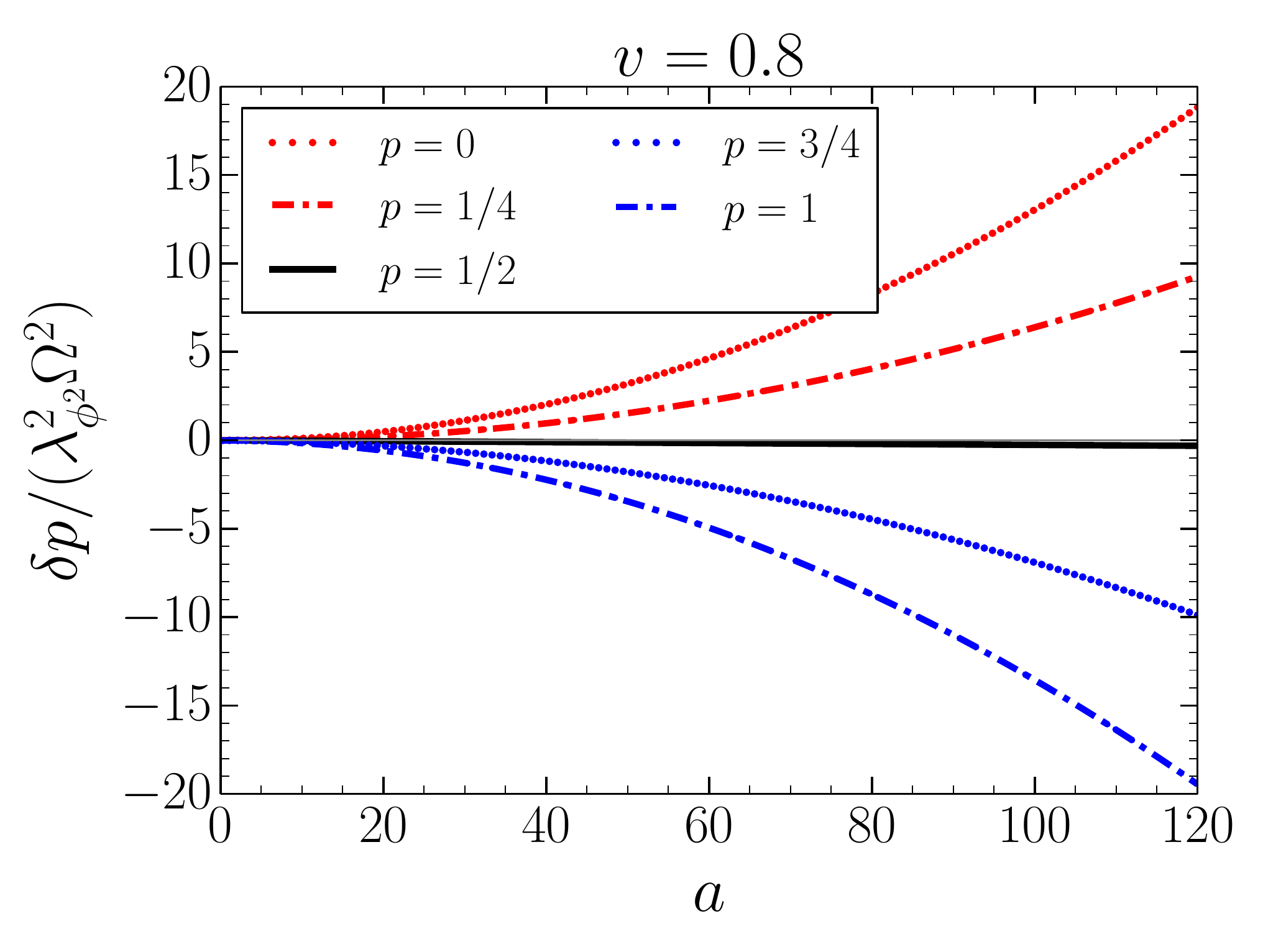}
            \caption{}
            \label{fig:v8QuadOmega}
        \end{subfigure}
        \caption{\small For fixed $v=0.8$, we show the behaviour of the population change for the quadratic scalar interaction, eq.~(\ref{eq:PopChange2}), for a range of initially excited probabilities, as a function of the reduced acceleration $a=\alpha/\Omega$. Figure (a) shows $\delta p_{\phi^2}$ in terms of  $\lambda^2_{\phi^2}/\mathcal{T}^2$ while (b) presents this in terms of $\lambda^2_{\phi^2}\Omega^2$}
        \label{fig:v8Quad}
    \end{figure}
\begin{figure}[h!]
        \centering
        \begin{subfigure}[b]{0.475\textwidth}
            \centering
            \includegraphics[width=\textwidth]{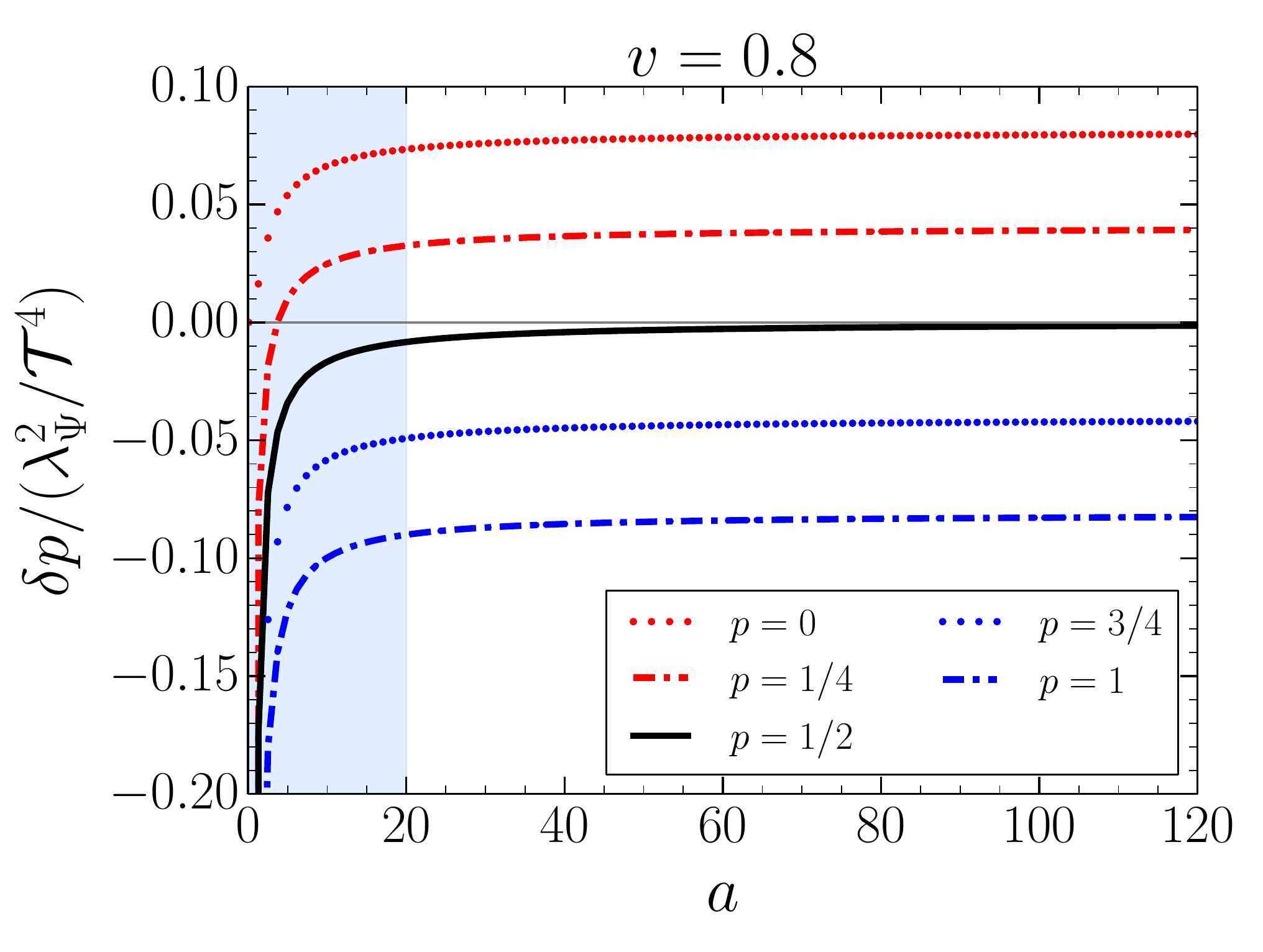}
            \caption{}
            \label{fig:v8FermT}
        \end{subfigure}
        \begin{subfigure}[b]{0.475\textwidth}  
            \centering 
            \includegraphics[width=\textwidth]{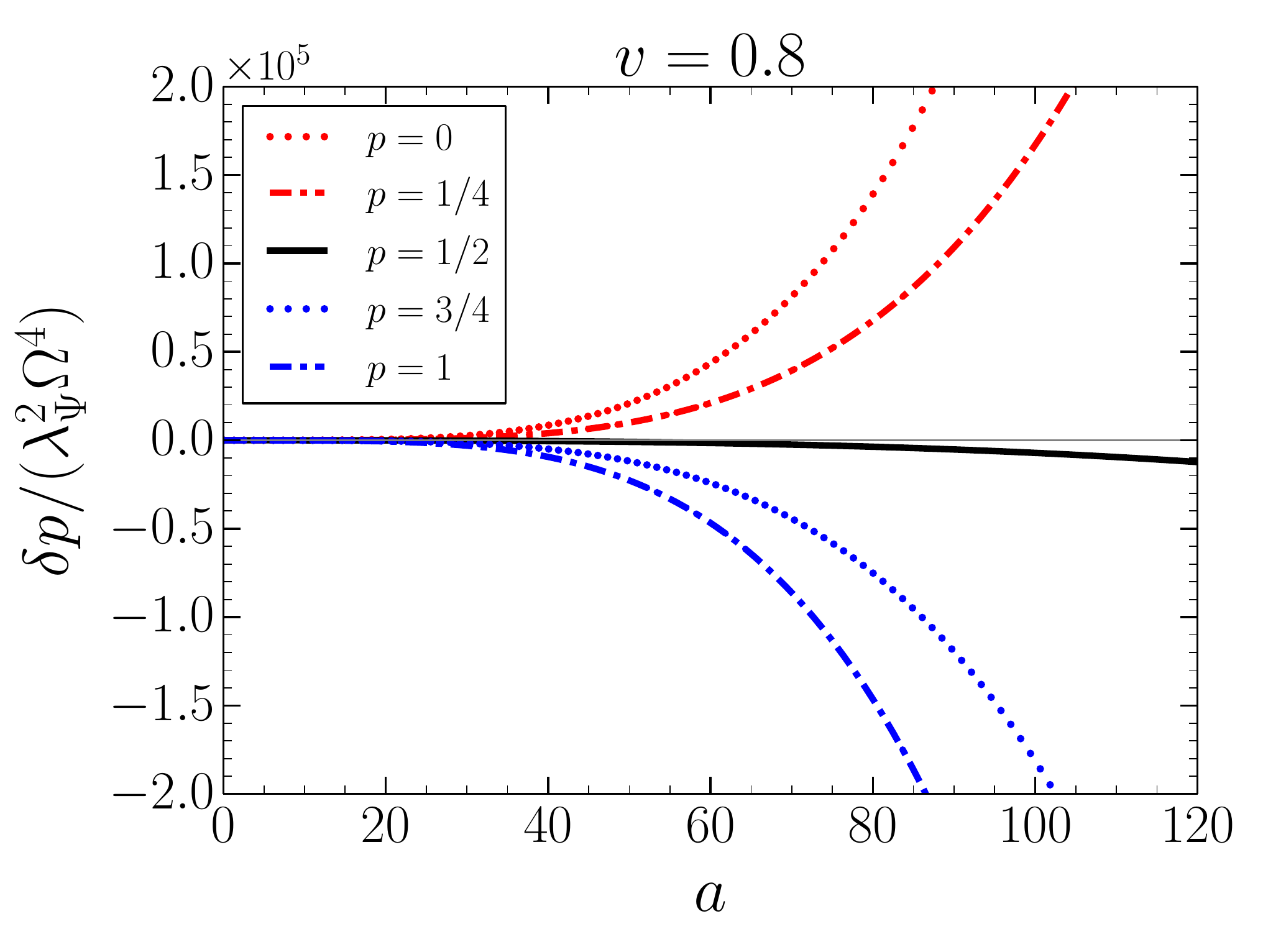}
            \caption{}
            \label{fig:v8FermOmega}
        \end{subfigure}
        \caption{ \small For fixed $v=0.8$, we plot the population change for the fermionic interaction, eq.~(\ref{eq:PopChange2}), for a range of initially excited probabilities. The left figure (a) shows $\delta p_{\Psi}$ in terms of  $\lambda^2_{\Psi}/\mathcal{T}^4$ while (b) presents this in terms of $\lambda^2_{\Psi}\Omega^4$.}
        \label{fig:v8Ferm}
    \end{figure}

 Let us now examine the behaviour of the population change, eq.~(\ref{eq:PopChange2}) in the quadratic and fermionic cases. In Figures~\ref{fig:v8QuadT} and \ref{fig:v8FermT} we plot $\delta p_F/(\lambda_F/\mathcal{T}^{\Delta_F})^2$ for a range of different initially excited probabilities and $v=0.8$ as in Figure~\ref{fig:p0Scalar4d}. This is a relevant way of presenting the population change when the timescale of interaction does not depend on the acceleration. In this case it is quite clear that both the quadratic and fermionic detectors respond in a manner that is qualitatively similar to the linear scalar case. Only the numerical scale changes, and so  all the previous comments on the linear scalar case apply.

 However for the kinematic steps in the Unruh quantum Otto cycle, the timescales of interaction $\mathcal{T}_{1/2}$ explicitly depend on the accelerations. Therefore, as discussed above, in Figures~\ref{fig:v8QuadOmega} and \ref{fig:v8FermOmega} we plot $\delta p_F/(\lambda_F\Omega^{\Delta_F})^2$. We can see from these that for large accelerations $|\delta p_F|$ continues to grow with $a$. This essentially follows from the arguments presented earlier. Further, from eq.~(\ref{eq:PertCon}), for the perturbative approach to hold, we must have $(\lambda_F\Omega^{\Delta_F})^2$ bounded by the maximum acceleration $a_{\text{max}}^{1-2\Delta_F}$ we wish to consider. This means that in these figures, the coupling constants are bounded by
	\begin{equation}\label{eq:CoupConst}
	\lambda_{\phi^2}^2\Omega^2\lesssim 10^{-2}\;,\;\lambda_{\Psi}^2\Omega^4\lesssim 10^{-6}
	\end{equation} 
Now we are in a position to examine the thermodynamics of the cycle.

\subsection{Completing the cycle}
The key steps of the Unruh quantum Otto cycle are when the accelerating qubit is in contact with the vacuum. In step 2 the qubit is in contact with the hot quantum vacuum for time $\mathcal{T}_2=2\text{arctanh}(v)/\alpha_H$ and the heat absorbed (see eq.~\ref{eq:Heat2}) is $\langle Q_2\rangle=\Omega_1\delta p_H$, where $\delta p_H$ is given by eq.~(\ref{eq:PopChange2}) evaluated for $\mathcal{T}_2$. For positive work we must impose that the qubit absorbs heat from the vacuum, i.e. we must require $\delta p_H>0$. In step 4, the qubit is in contact with the cold quantum vacuum during time $\mathcal{T}_1=2\text{arctanh}(v)/\alpha_C$ and the heat transferred (see eq.~\ref{eq:Heat4}) is $\langle Q_4\rangle=\Omega_2\delta p_C$. Here again $\delta p_C$ is given by eq.~(\ref{eq:PopChange2}) now evaluated for $\mathcal{T}_1$. 

Now as a reminder, steps 1 and 3 consist of adiabatic expansion and contraction of the energy gap. No heat is transferred, while the total work in these steps is, c.f.~eq.(\ref{eq:Work}), $\langle W_\text{tot}\rangle=\langle W_1\rangle+\langle W_3\rangle=-(\Omega_2-\Omega_1)\delta p_H$. Moreover the total heat is $\langle Q_\text{tot}\rangle=(\Omega_2-\Omega_1)\delta p_H$, and so we have kept the first law of thermodynamics $\langle W_\text{tot}\rangle+\langle Q_\text{tot}\rangle=0$. In order to arrive at these expressions we have imposed the cyclic condition
	\begin{equation}\label{eq:Cyclic}
	\delta p_H(a_H,p,v)+\delta p_C(a_C,p,v)=0\;.
	\end{equation}
We will soon use this to ensure we get positive work from the qubit, but first let us consider the efficiency of the cycle. Since changing the gap involves work being done on the qubit externally, it is useful to consider the work done by the qubit $\langle W_{\text{ext}}\rangle=-\langle W\rangle$. Therefore the standard efficiency is simply the ratio 
	\begin{equation}
	\eta=\frac{\langle W_{\text{ext}}\rangle}{Q_2}=1-\frac{\Omega_1}{\Omega_2} 
		\end{equation}
of total work to heat from the hot thermal bath. 
This only depends on the ratio of the two energy levels and does not involve the temperatures (accelerations) of the thermal bath. It also corresponds to the efficiency found in~\cite{Kieu2004} for the quantum Otto cycle with a standard thermal bath suggesting a universal bound that does not depend on the details of the reservoirs~\cite{Arias2018}. A more realistic measure of the efficiency of this cycle would be to put some bounds on the work required to accelerate the detectors throughout this cycle. However this is beyond the scope of this paper, which is focused on just an idealized proof of concept.

\begin{figure}[h!]
        \centering
        \begin{subfigure}[b]{0.475\textwidth}
            \centering
            \includegraphics[width=\textwidth]{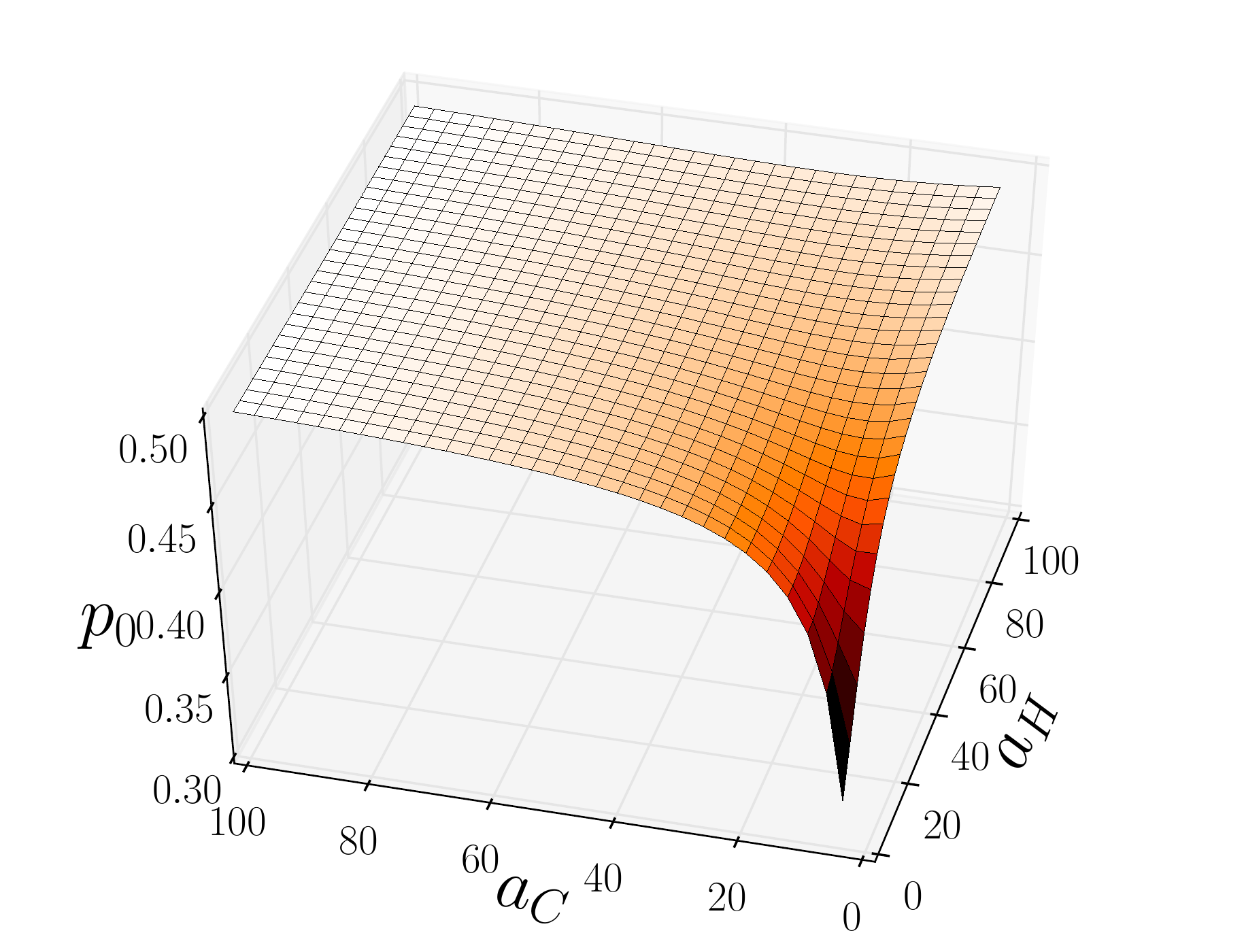}
            \caption{}
            \label{fig:p0plotQ}
        \end{subfigure}
        \begin{subfigure}[b]{0.475\textwidth}  
            \centering 
            \includegraphics[width=\textwidth]{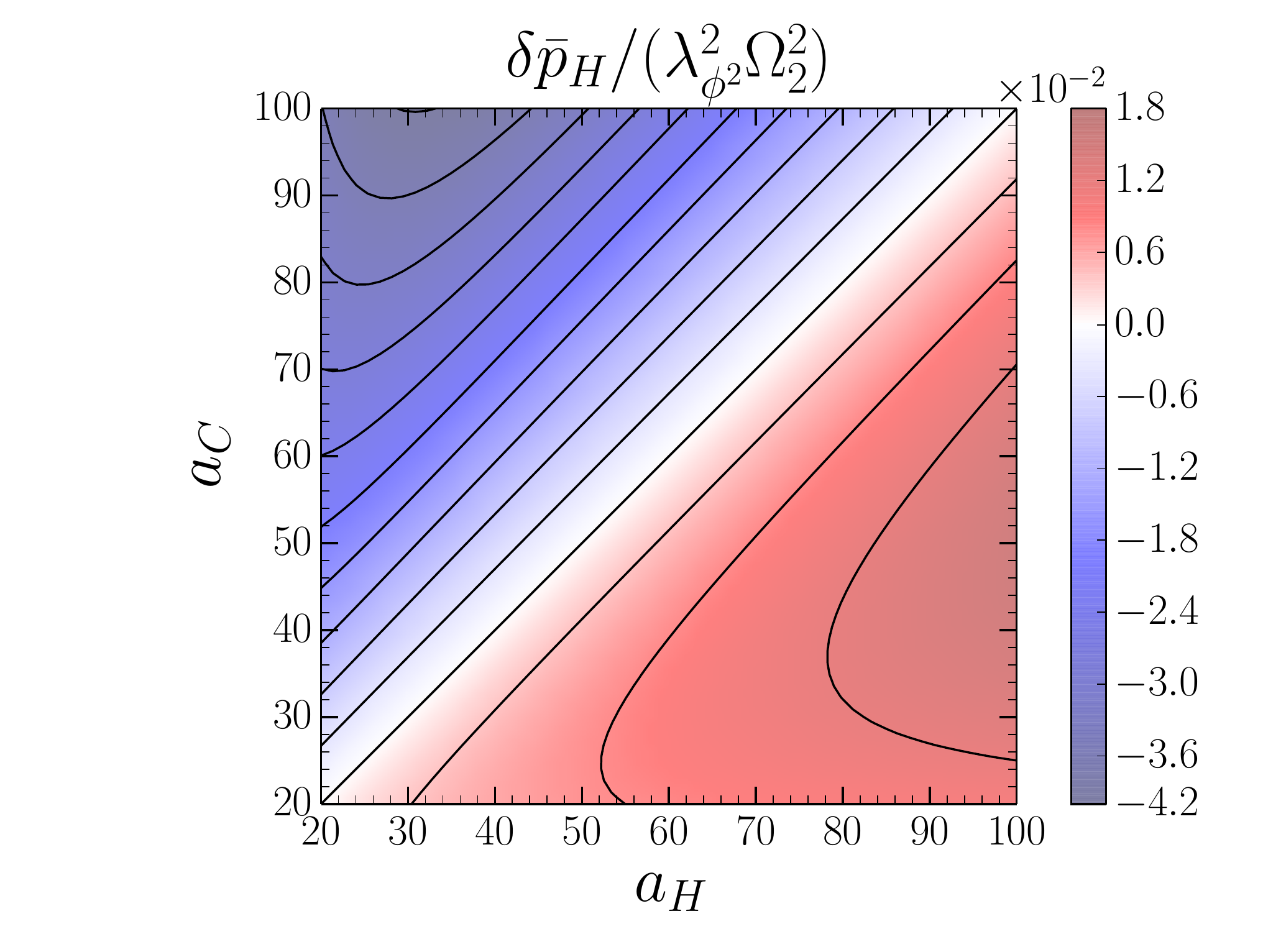}
            \caption{}
            \label{fig:WQ}
        \end{subfigure}
        \caption[]
        {\small For the quadratically coupled scalar interaction, with $v=0.8$ and $\Omega_2/\Omega_1=2$; figure (a) shows the behaviour of the critical probability for completing the thermodynamic cycle and (b) the correction $\delta\bar{p}_H$ as a function of the critical probability. Red regions show where we get positive work and blue negative.} 
        \label{fig:WPQuad}
    \end{figure}

\begin{figure}[h!]
        \centering
        \begin{subfigure}[b]{0.475\textwidth}
            \centering
            \includegraphics[width=\textwidth]{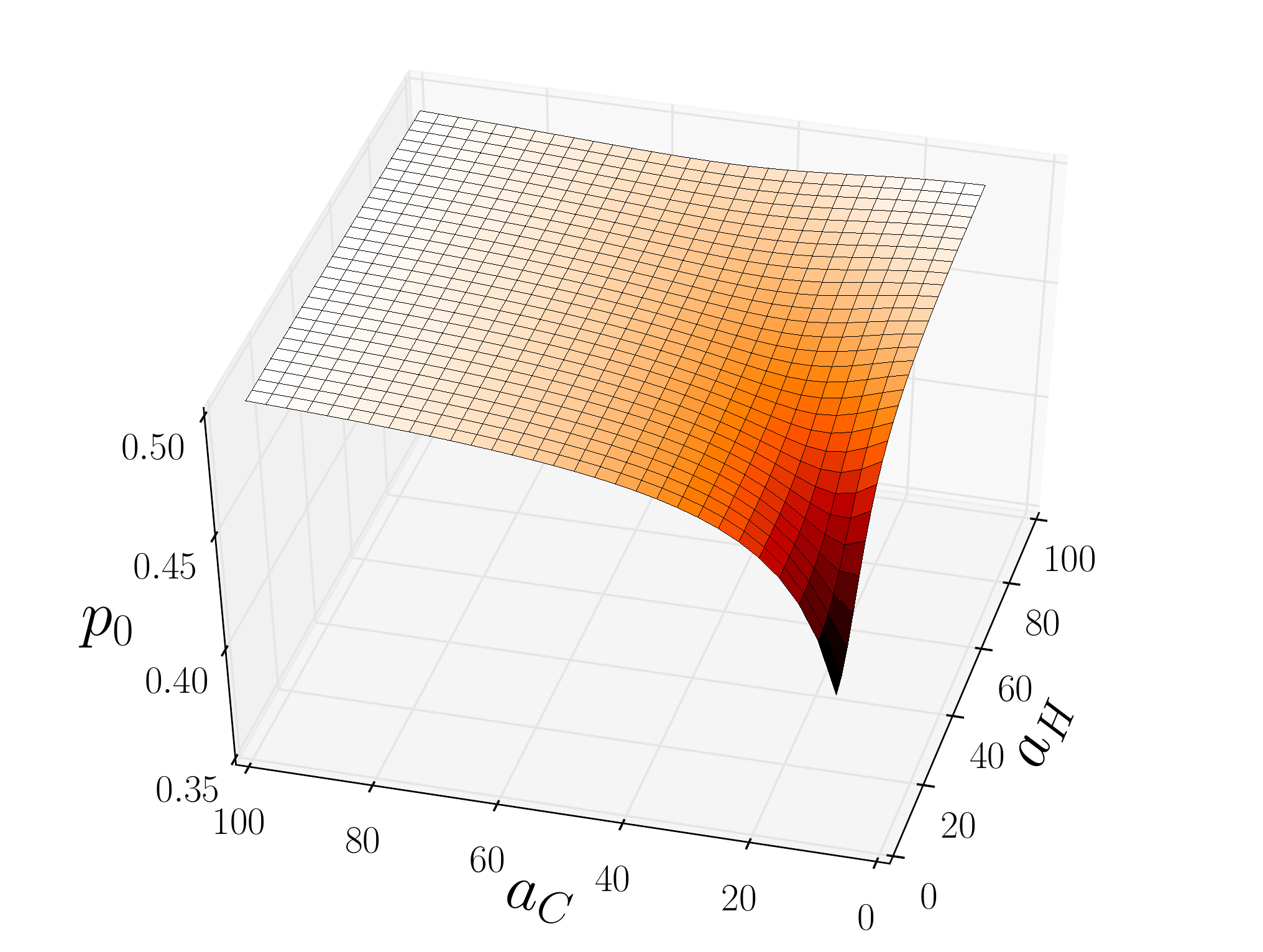}
            \caption{}
            \label{fig:p0plotF}
        \end{subfigure}
        \begin{subfigure}[b]{0.475\textwidth}  
            \centering 
            \includegraphics[width=\textwidth]{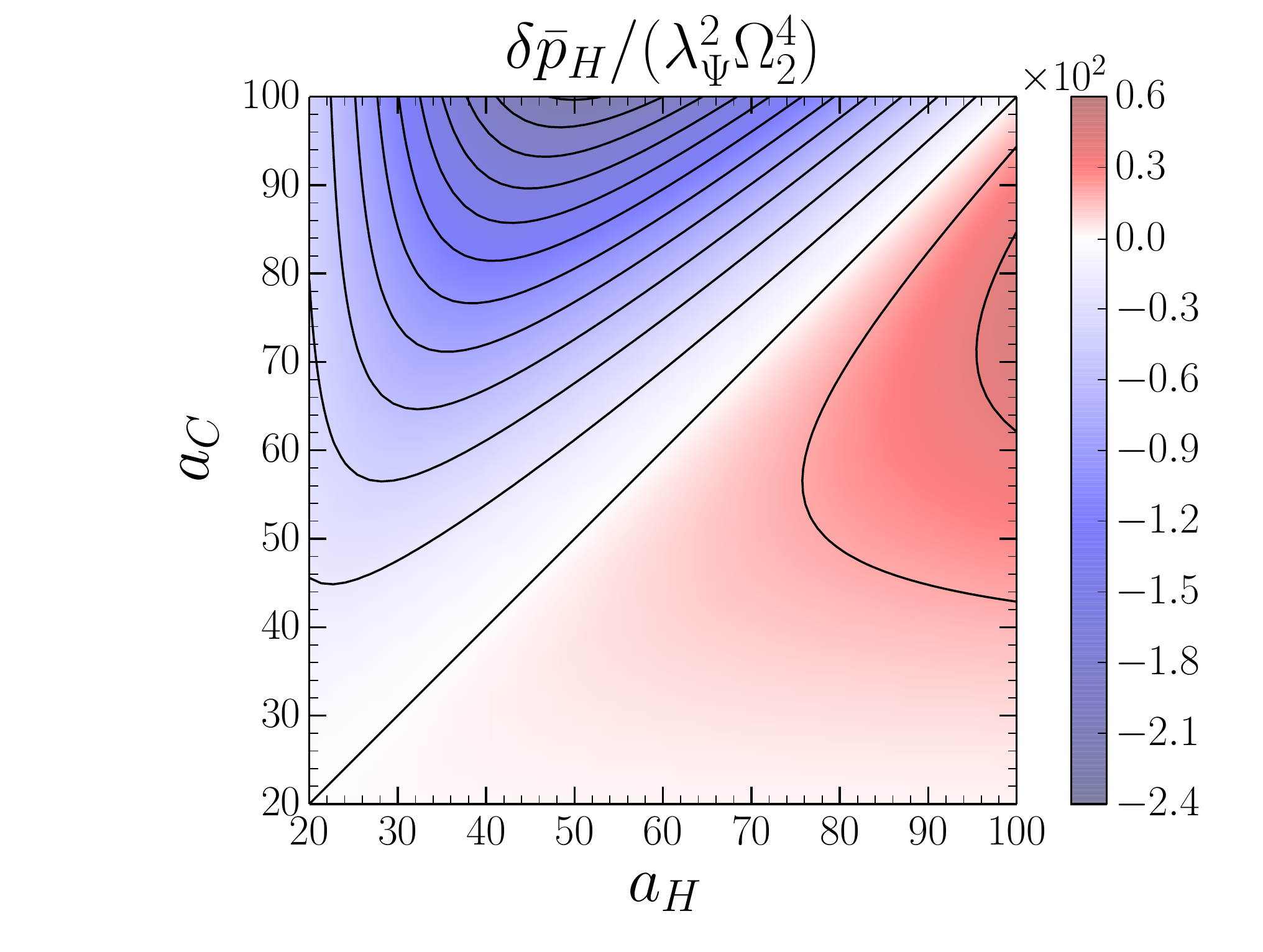}
            \caption{}
            \label{fig:WF}
        \end{subfigure}
        \caption[]
        {\small For the fermionic interaction with $v=0.8$ and $\Omega_2/\Omega_1=2$; (a) shows the critical probability for completing the thermodynamic cycle and (b) the correction $\delta\bar{p}_H$ as a function of the critical probability. Red regions show where we get positive work and blue negative.} 
        \label{fig:WPFerm}
    \end{figure}

Now we can use eq.~(\ref{eq:Cyclic}) as a condition on the initial excited population in order for the thermodynamic cycle to be complete. We can solve this for a critical probability for each coupling $p_0=p_0(a_H,a_C,v)$, such that for a given velocity $v$, hot acceleration $a_H$, and cold acceleration $a_H$, the vacuum always acts as a hot thermal bath in step 2 and a cold thermal bath during step 4. Using eq.~(\ref{eq:PopChange}) and the cyclic condition, $\delta p_H+\delta p_C=0$, it follows that
	\begin{equation}\label{eq:CritProb}
	p_0=p_0(a_H,a_C,v)=\frac{\mathcal{P}}{1+2\mathcal{P}}
	\end{equation}
where we have defined
	\begin{equation}
	\mathcal{P}=\frac{\mathcal{F}_F(1/a_H,2\text{arctanh}[v])+\mathcal{F}_F(1/a_C,2\text{arctanh}[v])}{\Delta\mathcal{F}_F(1/a_H,2\text{arctanh}[v])+\Delta\mathcal{F}_F(1/a_C,2\text{arctanh}[v])}\;.
	\end{equation}
 Here again the dimensionality enters into the problem for the quadratic and ferminonic cases. Since $\delta p_H/\delta p_C\propto(\mathcal{T}_1/\mathcal{T}_2)^{2\Delta_F}\propto(a_H/a_C[\Omega_2/\Omega_1])^{2\Delta_F}$, c.f. eq.~(\ref{eq:popdim}). This means that in these instances the gap ratio $\Omega_2/\Omega_1$ is now a parameter that  must be fixed in order to plot the behaviour of the thermodynamic quantities.
\begin{figure}[h!]
        \centering
        \begin{subfigure}[b]{0.475\textwidth}
            \centering
            \includegraphics[width=\textwidth]{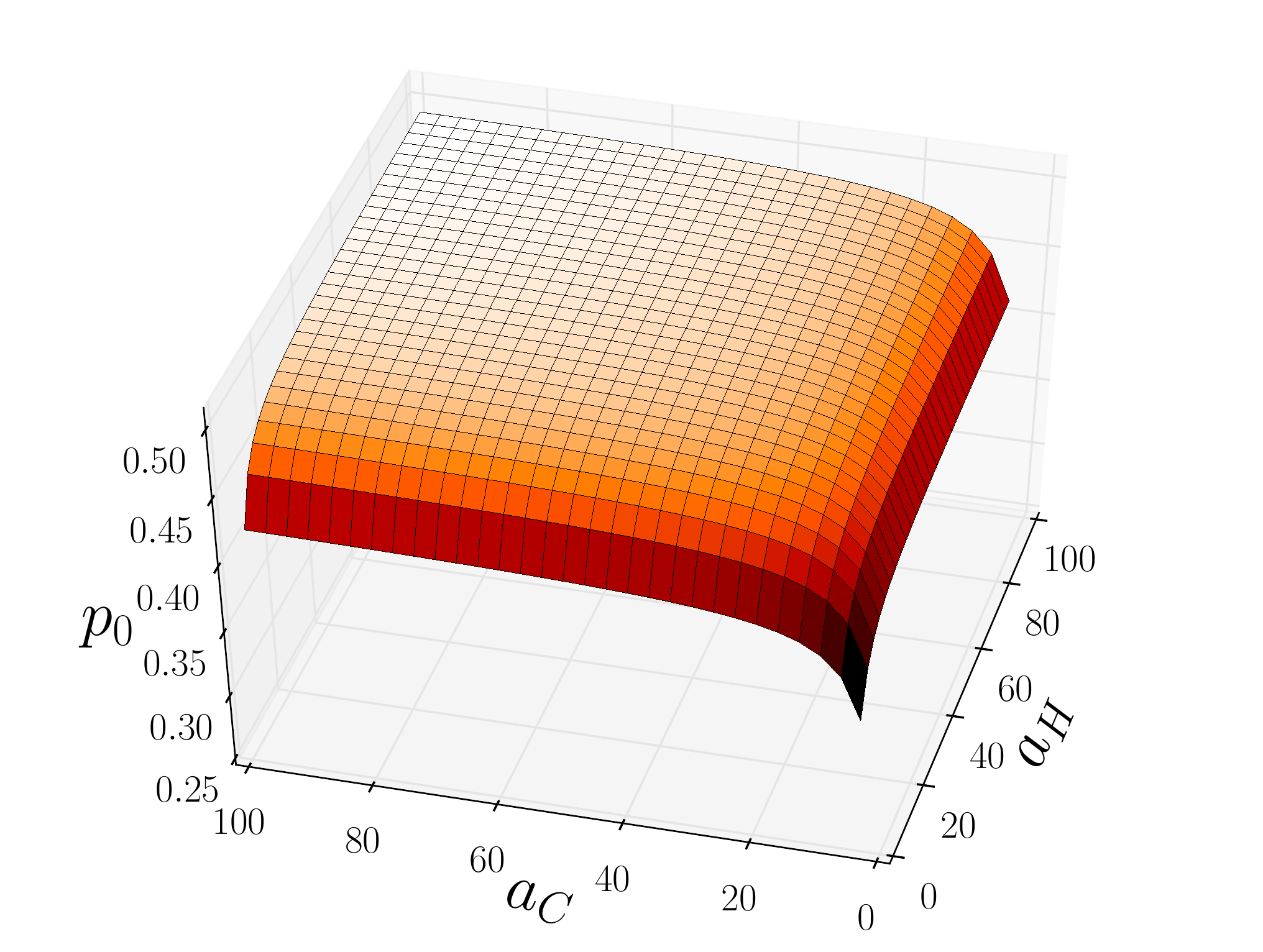}
            \caption{}
            \label{fig:p0plotS}
        \end{subfigure}
        \begin{subfigure}[b]{0.475\textwidth}  
            \centering 
            \includegraphics[width=\textwidth]{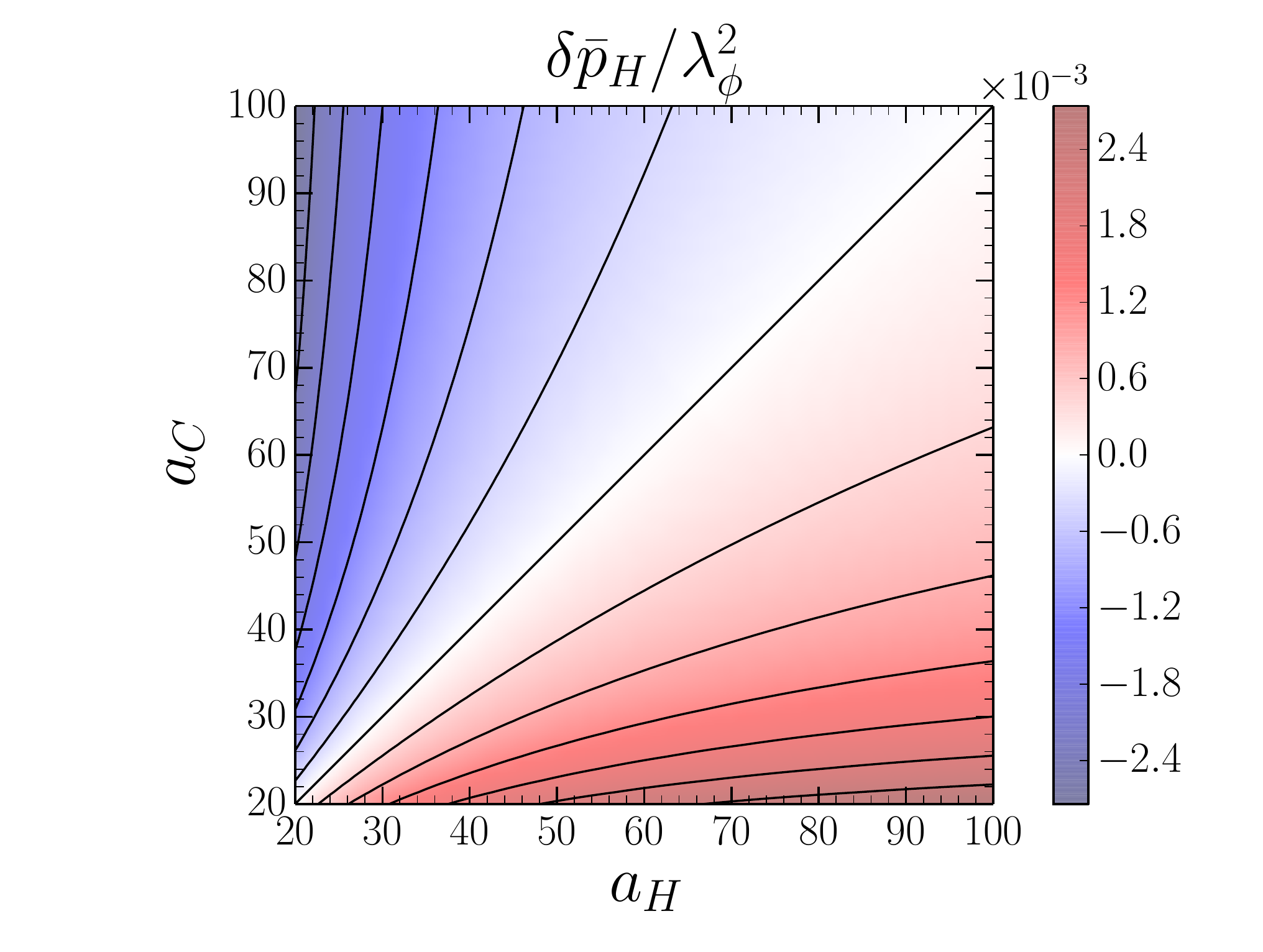}
            \caption{}
            \label{fig:WS}
        \end{subfigure}
        \caption{\small For $v=0.8$ and the linearly coupled scalar interaction: At left, (a) exhibits the nature of the critical probability for completing the thermodynamic cycle, and at right (b) plots the correction $\delta\bar{p}_H$ as a function of the critical probability. Red regions show where we get positive work and blue negative.}
        \label{fig:WPS}
\end{figure}
We plot $p_0$, for $v=0.8$, as a function of $a_H$ and $a_C$, for the linear scalar example, and with $\Omega_2/\Omega_1=2$ for the quadratic scalar and fermionic cases  in Figures~\ref{fig:p0plotS}, \ref{fig:p0plotQ}, \ref{fig:p0plotF} respectively. In these we can see that we always have $0\leq p_0<1/2$, as discussed above. Also as both $a_C$ and $a_H$ grow, $p_0$ approaches $\sim.49$.  However the quadratic and fermionic coupling exhibit qualitatively different behaviour compared to the  linear scalar case. They have a distinct asymmetry about the $a_H=a_C$ line. This arises from the dependence on the gap ratio and gets more pronounced as the ratio is increased.

We can now use $p_0$ to fix the correction for which the thermodynamic cycle is complete. That is to say, using eqs.~(\ref{eq:PopChange2}, \ref{eq:CritProb}) we define
	\begin{equation}\label{eq:Woutput}
	\delta\bar{p}_H=\delta\bar{p}_H(a_H,a_C,v)=\delta p_F(a_H,p_0,v)
	\end{equation}
Then the cyclic condition eq.~(\ref{eq:Cyclic}) is automatically satisfied with $\delta\bar{p}_C=-\delta\bar{p}_H$. This is plotted in Figures~\ref{fig:WS}, \ref{fig:WQ}, \ref{fig:WF} for each case, again, as a function of $a_H$ and $a_C$ with $v=0.8$ and $\Omega_2/\Omega_1=2$. In these figures regions where the correction $\delta\bar{p}_H$, and hence the work, is positive are shown in red and negative regions are in blue. It is quite clear that in all cases the region of positive work is defined by the condition $a_H>a_C$. Thus, as in Arias et al.~\cite{Arias2018}, we find the same condition (eq.~(\ref{eq:PWCon})) as for the quantum Otto engine in ref.~\cite{Kieu2004}, with the role of temperature replaced by acceleration. That is to say
	\begin{equation}\label{eq:TempCon}
	\alpha_H/\Omega_2>\alpha_C/\Omega_1\;
	\end{equation}
which is again a stronger condition than in the classical case. We note that for the linearly coupled case our result is slightly different numerically than that of ref.~\cite{Arias2018}. This stems primarily from the missing $1/2$ prefactor in eq.~(B14) of ref.~\cite{Arias2018}. 

While the linear scalar coupling reproduces the structure of the original quantum Otto cycle (see Figure~\ref{fig:WorkQO}), we find that the quadratic and fermionic cases exhibit qualitatively different behaviour. The asymmetry about $a_H=a_C$ is evident here as well, and is more pronounced in the fermionic case. We can also see that the transition between the regions of positive and negative $\delta \bar{p}_H$ are much sharper than for the linear coupling. Further, in the linear case and the ordinary quantum Otto cycle, the work is largest when the difference between the hot and cold accelerations/temperatures is largest. However, this no longer holds for quadratic and fermionic detectors, due to the extra dependence on the ratio $(a_H/a_C)^{2\Delta_F}$ arising from the dimensionality of the coupling constants/response functions. There is a subtle balancing act at play between the interaction timescales.

Ideally we would compare our work output from this cycle to the quantum Otto cycle in ref.~\cite{Kieu2004}, i.e. eq.~(\ref{eq:Wcl}). Certainly the response functions divided by the time interval, $\mathcal{F}_F(\Omega,\mathcal{T})/\mathcal{T}$, approach thermal KMS states based on the long time limit~\cite{Fewster2016,Garay2016} (see eq.~(\ref{eq:LongT}) and Appendix~\ref{App:LongTime}). However this means that the response function itself diverges linearly with $\mathcal{T}$~\cite{Fewster2016}, as $\mathcal{T}\rightarrow\infty$. Thus the population change eq.~(\ref{eq:PopChange2}) also diverges. To understand how the work of the Unruh quantum Otto cycle compares to that of an exactly thermal bath, one could perhaps use the bounds on the interaction time introduced in ref.~\cite{Fewster2016}. These quantify how close to thermal (KMS) equilibrium the response function of the UdW detector is, for a finite interaction time. Alternatively, a $1/\mathcal{T}$ expansion of the response function, as in ref.~\cite{Garay2016}, may useful to this end. These approaches might also allow for more insight into the asymmetric structure of the quadratic and fermionic cycles. Evidently though, a finite interaction time still allows this thermodynamic cycle to be constructed with the same efficiency and temperature/acceleration bounds as in the original case.

In summary, to run this Unruh quantum Otto engine for given velocity and hot and cold vacuums/accelerations, we must first tune the state of the qubit such that it has the initial probability $p_0$ from eq.~(\ref{eq:CritProb}) then put it through the steps in Section~\ref{sec:UQOC}.  We close this section by noting that if we were to operate the cycle in the negative regime then it would act as a refrigerator.

%

\section{Conclusions and Outlook}\label{sec:Conc}
 
We have shown that the Unruh effect can be used to exploit the quantum vacuum as a thermal heat bath for a thermodynamic cycle, analogous to the Otto heat engine, for a variety of detector-field couplings, generalizing the proposal of Arias et al.~\cite{Arias2018} to include quadratically coupled scalar and fermionic fields.  We analytically calculated the response function of the UdW detector for each of these cases in 4 spacetime dimensions. Interestingly this response function depends only on the Wightman function of a scalar field as found in refs.~\cite{Sachs2017,Louko2016Fermions}. Furthermore, our forms of the response function eqs.~(\ref{eq:QResp}, \ref{eq:FResp}) for the quadratic scalar and fermion are, as far as we are aware, unknown in the literature. 

One of the key findings is that the conditions for completion of the cycle depend on the dimensionality of the different couplings. This  causes the cycle to behave qualitatively differently for the quadratic and fermionic cases. In particular, there is a dependence on the gap ratio of the detector  that introduces an asymmetry in the work output. Moreover, commensurate with the results of~\cite{Arias2018}, the Unruh quantum Otto cycle seems to preserve the condition~(\ref{eq:TempCon}), on the temperatures of the thermal baths in order to extract positive work, suggesting a universal nature of the bound found in ref.~\cite{Kieu2004}.

There are several suggestive routes for extending and complementing our understanding of the Unruh quantum Otto cycle. First, it would be instructive to understand how different switching functions affect the process. Quantifying modifications due to  finite detector size  would also be useful.  To do either, however, would require numerical integration, and care must be taken to regularize to $i\epsilon$ prescription, which gets very complicated for higher dimensions~\cite{Hodgkinson2011}. This is essential for the quadratic and fermionic case, as  eqs.~(\ref{eq:QExp}, \ref{eq:FExp}) indicate. 

Second, one of the distant goals in the field is to experimentally verify the Unruh effect. With this in mind, it would be very useful to consider whether a more experimentally realizable kinematic trajectory for the detector can be used for the cycle and still keep the thermal quantum vacuum. It is crucial to determine both the input cost of this process and  the parameter range required to get an experimentally detectable output.  

It would also be of particular interest to understand how long the system takes to thermalize. This could perhaps be done using the bounds in ref.~\cite{Fewster2016} or by developing a $1/\mathcal{T}$ expansion as in ref.~\cite{Garay2016}, using the asymptotic forms of the Lerch--Hurwitz transcendent~\cite{Ferreira2004}. Finally, from a gravitational perspective it would be intriguing to see how this cycle could be generalized to curved spacetimes where we use black hole thermodynamics and the Hawking temperature. We reserve these questions for   future explorations.

\appendix

\section{Quantum field theory conventions}\label{App:QFT Con}

In this appendix we present our conventions for the massless scalar and fermionic fields (see refs.~\cite{Birrell1984quantum,Peskin1995QFT} for more details). 
\subsection{Scalar fields}

We consider a massless scalar field, $\phi(x)$ with Lagrangian density,
	\begin{equation}\label{eq:Lag}
	\mathcal{L}=\frac{1}{2}\partial^a\phi\partial_a\phi\;,
	\end{equation}
with canonical conjugate momentum $\pi(x)=\frac{\partial\mathcal{L}}{\partial(\partial_0\phi)}=\partial_0\phi(x)$.
This allows the Hamiltonian  to be constructed via the Legendre transform,
	\begin{align}\label{eq:Ham}
	H_\phi&=\int \dd[{d-1}]{x}\;\frac{1}{2}\left[(\partial_t \phi)^2+(\nabla\phi)^2\right]\;.
	\end{align}
Variation of the action leads to the Klein-Gordon equation
	\begin{equation}\label{eq:KG}
	\square\phi(x)=(\partial^2_t-\nabla^2)\phi(x)=0\;,
	\end{equation}
which has plane wave solutions, 
	\begin{equation}\label{eq:Modes}
	u_{\mathbf{k}}(x)=\frac{1}{\sqrt{2\omega}}\;e^{-ikx}\;,
	\end{equation}
where $kx=k_\mu x^\mu=k^0\;x^0-\vb{k}\cdot\vb{x}$ and $k^0=|\mathbf{k}|$. These are called positive frequency modes with respect to $t$. One can then define the inner product,
	\begin{equation}\label{eq:IP}
	(\phi_1,\phi_2)=-i\int \dd[d-1]{x}\;\Big( \phi_1(x)\partial_t\phi^*_2(x)-\partial_t[\phi_1(x)]\phi^*_2(x)\Big)\;.
	\end{equation}
It can be seen that the positive frequency modes eq.~(\ref{eq:Modes}) form a complete orthogonal basis and so the field can be expanded
	\begin{equation}\label{eq:PhiExpan}
	\phi(t,\mathbf{x})=\int \widetilde{\dd k}\;\left( a_{\mathbf{k}}u_k(x)+a^\dagger_{\mathbf{k}}u^*_k(x)\right)\;.
	\end{equation}
where $\widetilde{\dd k}=\dd[d-1]{k}/((2\pi)^{d-1}2k^0)$ is the Lorentz invariant phase space measure. We employ the standard quantisation scheme and impose the equal time commutation relations,
	\begin{align}\label{eq:Comm1}
	\left[\phi(t,\mathbf{x}),\phi(t,\mathbf{x}') \right]=0\;,\;
	\left[\pi(t,\mathbf{x}),\pi(t,\mathbf{x}') \right]=0\;\;
	\left[\phi(t,\mathbf{x}),\pi(t,\mathbf{x}') \right]=i\;\delta^{(d-1)}(\mathbf{x}-\mathbf{x'})\;,
	\end{align}
Using~(\ref{eq:PhiExpan}), the commutation relations are equivalent to imposing
	\begin{align}\label{eq:Comm2}
	\big[a_{\mathbf{k}},a_{\mathbf{k}'}\big]=0\;,\;
	\big[a^\dagger_{\mathbf{k}},a^\dagger_{\mathbf{k}'}\big]=0\;,\;
	\big[a_{\mathbf{k}},a^\dagger_{\mathbf{k}'}\big]=2k^0(2\pi)^{d-1}\delta^{(d-1)}(\mathbf{k}-\mathbf{k}') 
	\end{align}
and so $a^{(\dagger)}_{\mathbf{k}}$ are the annihilation (creation)  operators.  The Fock-Space representation of the Hilbert space is constructed from the vacuum state $\ket{0}$ defined by
	\begin{equation}\label{eq:Vac}
	a_{\mathbf{k}}\ket{0}=0=\bra{0}a^\dagger_{\mathbf{k}}\;.
	\end{equation}
The vacuum is Lorentz invariant and contains no particles. Furthermore the operator $a^\dagger_{\mathbf{k}}$ creates a particle in the state $\mathbf{k}$, $\ket{1_{\mathbf{k}}}=a^\dagger_{\mathbf{k}}\ket{0}$ and many particle states can be built with repeated application of $a^\dagger_{\mathbf{k}}$.

\subsection{Fermionic fields}

Following essentially the conventions of ref.~\cite{Louko2016Fermions} the massless fermionic Lagrangian density is given by
	\begin{equation}
	\mathcal{L}=\overline{\Psi}\;i\gamma^\mu\partial_\mu\Psi
	\end{equation}
where as usual the conjugate spinor is  $\overline{\Psi}=\Psi^\dagger\gamma^0$ and the gamma matrices are $N_d\times N_d$ dimensional matrices where
	\begin{equation}
	N_d=\begin{cases}
	2^{d/2}&d \text{ even}\\
	2^{(d-1)/2}&d \text{ odd}\\
	\end{cases}
	\end{equation} 
which satisfy the Clifford algebra
	\begin{equation}
	\{\gamma^\mu,\gamma^\nu\}=2\eta^{\mu\nu}\mathds{1}_{N_d\times N_d}\;.
	\end{equation}
Moreover $\gamma^0$ is Hermitian and $\gamma^1,...,\gamma^{d-1}$ are anti-Hermitian. From the Lagragian we have canonical momentum $\Pi=\Psi^\dagger$ and so the Hamiltonian of the massless Dirac field is given by
	\begin{equation}
	H_\Psi=\int\dd[d-1]{x}\overline{\Psi}\gamma^j\partial_j\Psi\;,\;j\in\{1,2,3\}\;.
	\end{equation}
Variation of the action leads to the massless Dirac equation for the spinor $\Psi$
	\begin{equation}
	i\gamma^\mu\partial_\mu\Psi=0
	\end{equation}
with a similar equation for its conjugate. We can form a complete set of solutions with the spinors
	\begin{subequations}
	 	\begin{alignat}{2}
		u^{(\varsigma)}_{\mathbf{k}}=& u^{(\varsigma)}(\mathbf{k})e^{-i kx}\\
		v^{(\varsigma)}_{\mathbf{k}}=& v^{(\varsigma)}(\mathbf{k})e^{+ikx}
		\end{alignat}
	\end{subequations}
where $k^0=|\mathbf{k}|$ and $\varsigma$ labels the spin polarization. The spinors $u^{(\varsigma)}(\mathbf{k})$, $v^{(\varsigma)}(\mathbf{k})$ satisfy the following properties;
	\begin{subequations}
	 	\begin{alignat}{4}
	 	\gamma^\mu k_\mu\; u^{(\varsigma)}(\mathbf{k})&=0=\gamma^\mu k_\mu\; v^{(\varsigma)}(\mathbf{k})\;,\\
		u^{(\varsigma)\dagger}(\mathbf{k})\;u^{(\varsigma')}(\mathbf{k}')&=+v^{(\varsigma)\dagger}(\mathbf{k})\;v^{(\varsigma')}(\mathbf{k}')= 2k^0\delta^{\varsigma\varsigma'}\;,\\
		\bar{u}^{(\varsigma)}(\mathbf{k})\;u^{(\varsigma')}(\mathbf{k}')&=-\bar{v}^{(\varsigma)}(\mathbf{k})\;v^{(\varsigma')}(\mathbf{k}')= 0\;,\\
		\bar{u}^{(\varsigma)}(\mathbf{k})\;v^{(\varsigma')}(\mathbf{k})&=0\;,
		\end{alignat}
	\end{subequations}
and
	\begin{equation}\label{eq:SpinSums}
	\sum_\varsigma\;u^{(\varsigma)}_a(\mathbf{k})\bar{u}^{(\varsigma)}_b(\mathbf{k}')=(\gamma^\mu k_\mu)_{ab}=\sum_\varsigma\;v^{(\varsigma)}_a(\mathbf{k})\bar{v}^{(\varsigma)}_b(\mathbf{k}')\;.
	\end{equation}
Here $a$ and $b$ explicitly label the spinor indices. The inner product between two spinors is defined as
	\begin{equation}
	(\Psi,\Phi)=\int \dd[d-1]{x}\;\overline{\Psi}(t,\mathbf{x})\Phi(t,\mathbf{x})\;.
	\end{equation}
It follows from the above expressions that the mode spinors satisfy the orthogonality relations
	\begin{subequations}
		\begin{alignat}{2}
		(u^{(\varsigma)}_{\mathbf{k}},u^{(\varsigma')}_{\mathbf{k'}})&=(v^{(\varsigma)}_{\mathbf{k}},v^{(\varsigma')}_{\mathbf{k'}})=2k^0(2\pi)^{d-1}\delta^{\varsigma\varsigma'}\delta^{(d-1)}(\mathbf{k}-\mathbf{k}')\;,\\
		(u^{(\varsigma)}_{\mathbf{k}},v^{(\varsigma')}_{\mathbf{k'}})&=0\;.
		\end{alignat}
	\end{subequations}
As in the scalar field case we can expand the field in the mode functions
	\begin{equation}
	\Psi(x)=\int \widetilde{\dd k}\;\sum_\varsigma\left(b_\varsigma(\mathbf{k})\;u^{(\varsigma)}_{\mathbf{k}}(x)+d^\dagger_\varsigma(\mathbf{k})\;v^{(\varsigma)}_{\mathbf{k}}(x)\right)\;,
	\end{equation}
and quantisation is implemented the standard way by imposing equal time anti-commutation relations on the field
	\begin{subequations}
		\begin{alignat}{2}
		\left\{\Psi_a(t,\mathbf{x}),\Psi_b(t,\mathbf{y})\right\}&=\left\{\Psi_a^\dagger(t,\mathbf{x}),\Psi_b^\dagger(t,\mathbf{y})\right\}=0\;,\\
		\left\{\Psi_a(t,\mathbf{x}),\Psi_b^\dagger(t,\mathbf{y})\right\}&=\delta_{ab}\;\delta^{(d-1)}(\mathbf{x}-\mathbf{y})\;.
		\end{alignat}
	\end{subequations}
Or equivalently the only non zero commutators of the expansion coefficients are
	\begin{align}\label{eq:AntiComm}
&		\left\{b_\varsigma(\mathbf{k}),b^\dagger_{\varsigma'}(\mathbf{k}')\right\} =	\left\{d_\varsigma(\mathbf{k}),d^\dagger_{\varsigma'}(\mathbf{k}')\right\}= 2k^0(2\pi)^{d-1}\delta_{\varsigma\varsigma'}\delta^{(d-1)}(\mathbf{k}-\mathbf{k}')\;.
	\end{align}
Again we construct the Fock space representation by defining the now fermionic vacuum state $\ket{0}$
	\begin{subequations}
		\begin{alignat}{2}
			b_\varsigma(\mathbf{k})\ket{0}&=0=\bra{0}b^\dagger_\varsigma(\mathbf{k})\;,\\
			d_\varsigma(\mathbf{k})\ket{0}&=0=\bra{0}d^\dagger_\varsigma(\mathbf{k})\;.
		\end{alignat}
	\end{subequations}
Excitations of the field are given by application of creation operators to this state. By virtue of the anti-commutation relations these states are antisymmetric under particle exchange and can only have occupation number one or zero. Thus Fermi--Dirac statistics are accounted for. 

\section{Vacuum correlation functions}\label{App:VCF}

In this appendix we calculate the vacuum correlators that appear in the response functions for the various Unruh--DeWitt detector models. In particular we shall see that, for massless fields, all of these correlators depend only on the scalar Wightman function.


The relevant vacuum correlation function for the response function of the linearly coupled UdW detector is just the scalar Wightman function defined as 
	\begin{equation}
	\mathcal{W}_\phi(x,y)=\bra{0}\phi(x)\phi(y)\ket{0}
	\end{equation}
Using the mode expansion eq.~(\ref{eq:PhiExpan}) and the properties of the scalar creation/annihilation operators eqs.~(\ref{eq:Comm2},\ref{eq:Vac}) this becomes
	\begin{align}\label{eq:WightCalc}
	\mathcal{W}_\phi(x,y)&=\int \widetilde{\dd k}\int \widetilde{\dd k'}\;2k^0(2\pi)^{d-1}\delta^{(d-1)}(\mathbf{k}-\mathbf{k}')\;e^{-ikx}e^{+ik'y}\nonumber\\ 
	&=\int \widetilde{\dd k}\;e^{-ik(x-y)-k^0\epsilon}\nonumber\\
	&=\int \frac{\dd[d]{k}}{(2\pi)^{d-1}}\Theta(k^0)\delta(k^2)\;e^{-ik(x-y)-k^0\epsilon}\;
	\end{align}
where in the second to last line we have explicitly included the regulator $\epsilon>0$ and the Wightman function must be understood as a distribution in the limit $\epsilon\rightarrow0$. In the last line we have written the measure $\widetilde{\dd k}=\dd[d-1]{k}/((2\pi)^{d-1}2k^0)$ in an explicitly Lorentz invariant manner. 

Defining $\Delta=|(x-y)|$ and employing a Lorentz transformation, we obtain~\cite{Takagi1986} 
	\begin{align}
		\mathcal{W}_\phi(x,y)&=\int \frac{\dd[d]{\bar{k}}}{2\pi^{d-1}}\Theta(\bar{k}^0)\delta(\bar{k}^2)\;e^{-i\bar{k}^0\Delta\text{sgn}(x^0-y^0)-\bar{k}^0\epsilon}\nonumber\\
		&=\int \widetilde{\dd \bar{k}}\;e^{-|\bar{k}|z}\;,\;\text{for }z\equiv\epsilon+i\Delta\text{sgn}(x^0-y^0)\nonumber\\
		&=\frac{1}{2(2\pi)^{d-1}}\int\dd\Omega_{d-1}\int_0^\infty \dd{k} k^{d-3}e^{-k z} 
	\end{align}
where $\bar{k}^0=|\bar{k}|\equiv k$. Performing the last integrations over the angular part and $k$ gives
	\begin{equation}\label{eq:SWightA}
	\mathcal{W}_\phi(x,y)=\frac{\Gamma(d/2-1)}{4\pi^{d/2}}[z(x,y)]^{2-d}\;,
	\end{equation}
where $\Gamma(z)\equiv\int_0^\infty t^{z-1}e^{-t}\dd{t}$ is the usual gamma function.

For the quadratically coupled detector \cite{Sachs2017,Hummer2016} the response function depends on the vacuum correlation
	\begin{equation}
		\mathcal{W}_{\phi^2}(x,y)=\bra{0}:\phi^2(x):\;:\phi^2(y):\ket{0}  =\bra{0}\phi_x^+ \phi_x^+ \phi_y^- \phi_y^- \ket{0} =2[\phi_x^+,\phi_y^-]^2
	\end{equation}
where $\phi_x=\phi(x)$ and we have written the field 
	\begin{equation}
	\phi^+(x)=\int\widetilde{\dd k}\;a_{\vb{k}}u_{\vb{k}}(x)\;,\;\phi^-(x)=\int\widetilde{\dd k}\;a^\dagger_{\vb{k}}u^*_{\vb{k}}(x)
	\end{equation}
in terms of positive and negative frequency components $\phi=\phi^++\phi^-$. The above result follows upon
noting that 
	\begin{align}
&		:\phi^2(x): =\phi^2(x)-\bra{0}\phi^2(x)\ket{0} =(\phi_x^++\phi_x^-)^2-[\phi_x^+,\phi_x^-] =\phi_x^+\phi_x^++2\phi_x^-\phi_x^++\phi_x^-\phi_x^-\;.
	\end{align}	
Since $[\phi_x^+,\phi_y^-]=\bra{0}\phi_x^+\phi_y^-\ket{0}=\bra{0}\phi(x)\phi(y)\ket{0}$ we have the result
	\begin{equation}\label{eq:QuadWight}
	\mathcal{W}_{\phi^2}(x,y)=2[\mathcal{W}_\phi(x,y)]^2\;.
	\end{equation}
%
 
For the fermionic field~\cite{Louko2016Fermions} there are two Wightman functions we need to consider for the detector response function. Decomposing the field into positive and negative frequency components $\Psi(x)=\Psi^+(x)+\Psi^-(x)$,
\be 
		\Psi^+(x) =\int \widetilde{\dd k}\;\sum_\varsigma b_\varsigma(\mathbf{k})\;u^{(\varsigma)}_{\mathbf{k}}(x) 
	\qquad 	\Psi^-(x) =\int \widetilde{\dd k}\;\sum_\varsigma d^\dagger_\varsigma(\mathbf{k})\;v^{(\varsigma)}_{\mathbf{k}}(x) 
\ee 
the Wightman functions of the fermionic field are
	\begin{subequations}\label{eq:FWight}
		\begin{alignat}{2}
			S^+_{ab}(x,y)&=\bra{0}\Psi_a(x)\overline{\Psi}_b(y)\ket{0}=\left\{\Psi^+_a(x), \overline{\Psi}^-_b(y)\right\} =i\gamma^\mu\partial_{x^\mu}\mathcal{W}_\phi(x,y)\;,\\
			S^-_{ab}(x,y)&=\bra{0}\overline{\Psi}_b(y)\Psi_a(x)\ket{0}=\left\{\Psi^-_a(x), \overline{\Psi}^+_b(y)\right\}=-i\gamma^\mu\partial_{x^\mu}\mathcal{W}_\phi(y,x)\;.
		\end{alignat}
	\end{subequations}
  Employing the normal ordered operator
	\begin{align}
		:\overline{\Psi}(x)\Psi(x):&=\overline{\Psi}(x)\Psi(x)-\bra{0}\overline{\Psi}(x)\Psi(x)\ket{0}\nonumber\\
		&=\overline{\Psi}_a^+(x)\Psi_a^+(x)+\overline{\Psi}_a^-(x)\Psi_a^-(x)+\overline{\Psi}_a^-(x)\Psi_a^+(x)-\Psi_a^-(x)\overline{\Psi}_a^+(x)\;.
	\end{align}
we obtain
	\begin{align}
	\mathcal{W}_\Psi(x,y)&=\bra{0}:\overline{\Psi}_a\Psi_a(x):\;:\overline{\Psi}_b\Psi_b(y):\ket{0}\nonumber\\
	&=\bra{0}\overline{\Psi}_a^+(x)\Psi_a^+(x)\overline{\Psi}_b^-(y)\Psi_b^-(y)\ket{0}\nonumber\\
	&=\bra{0}\{\overline{\Psi}_a^+(x),\Psi_b^-(y)\}\{\Psi_a^+(x),\overline{\Psi}_b^-(y)\}\ket{0}\nonumber\\
	&=S_{ba}^-(y,x)\;S^+_{ab}(x,y)\nonumber\\
	&=\Tr\left[S^+(x,y)S^-(y,x)\right] \nonumber\\
	&=N_d\;\partial_{x^\mu}\mathcal{W}_\phi(x,y)\;\partial^{x^\mu}\mathcal{W}_\phi(x,y)
	\end{align}
This simplifies to
	\begin{align}
	\mathcal{W}_\Psi(x,y)&=N_d\;\partial_{x^\mu}\mathcal{W}_\phi(x,y)\;\partial^{x^\mu}\mathcal{W}_\phi(x,y)\nonumber\\
						 & =N_d\left(\frac{\Gamma(d/2-1)}{4\pi^{d/2}}\right)^2[\partial_{x^\mu}z(x,y)] [\partial^{x^\mu}z(x,y)]\left(\frac{d}{dz}z^{2-d}\right)^2\nonumber\\
						 &=N_d\left(\frac{(2-d)\Gamma(d/2-1)}{4\pi^{d/2}}\right)^2z^{2-2d}\;[\partial_{x^\mu}z(x,y)] [\partial^{x^\mu}z(x,y)]
	\end{align}
and upon using $z(x,y)=\epsilon+\text{sgn}(x^0-y^0)\Delta(x,y)$ and $\Delta(x,y)=\sqrt{(x^\mu-y^\mu)(x_\mu-y_\mu)}$ we have
	\begin{align}
& [\partial_{x^\mu}z(x,y)] [\partial^{x^\mu}z(x,y)] =\text{sgn}(x^0-y^0)^2[\partial_{x^\mu}\Delta(x,y)][\partial^{x^\mu}\Delta(x,y)] =\frac{(x_\mu-y_\mu)(x^\mu-y^\mu)}{\Delta^2} = 1	
\end{align}
where we have taken the limit $\epsilon\rightarrow0$.Thus we find the result
	\begin{equation}\label{eq:FExpA}
		\mathcal{W}_\Psi(x,y)=\frac{N_d\Gamma(d/2)^2}{4\pi^{d}}z^{2-2d}=\frac{N_d\Gamma(d/2)^2}{\Gamma(d-1)}\mathcal{W}_\phi^{(2d)}(x,y)
	\end{equation}
where $\mathcal{W}_\phi^{(2d)}(x,y)$ is the Wightman function of the massless scalar field in $2d$ dimensions. 
In the massive case the result above does not hold in general but can be written in terms of Bessel functions~\cite{Louko2016Fermions} (noting that ref.~\cite{Louko2016Fermions} does not consider normal ordered operators and so finds extra terms which diverge in the massive case).

\section{Evaluating the response function}\label{App:EvalResp}

The crux of the thermodymamic cycle lies in evaluating the change in the population of energy levels of the qubit. Whilst the form of the population change, eq.~(\ref{eq:Rhot}), is quite general, the details will change depending on  the dimensionality and the type of field. Now the dynamics of the population change are captured by
	\begin{align}\label{eq:Response}
	\mathcal{F}_F(\Omega,\mathcal{T})&=\frac{1}{2}\int_{-\infty}^{+\infty} \dd{s}\left(\mathcal{W}_F( s )e^{-i\Omega s } \int_{-\infty}^{+\infty} \dd{u}\chi_\mathcal{T}((u+s)/2)\chi_\mathcal{T}((u-s)/2)\right)\;,
	\end{align}
where we have changed variables to $ s =\tau-\tau'$ and $u=\tau+\tau'$. Following Arias et al.~\cite{Arias2018} we will consider a Lorentzian switching function
	\begin{equation}\label{eq:Lorentz}
	\chi_\mathcal{T}(\tau)=\frac{(\mathcal{T}/2)^2}{\tau^2+(\mathcal{T}/2)^2}\;.
	\end{equation}	
The advantage of the Lorentzian regulator is that it will enable us to extend the integration to the complex plane and use the residue theorem~\cite{Freitag2005} since $\chi_\mathcal{T}(z)\rightarrow0$ for $|z|\rightarrow\infty$ for all $z\in\mathbb{C}$. Note that if we had used a Gaussian switching function   this would fail.

The only $u$ dependence is in the switching functions so we can first evaluate
	\begin{align}\label{eq:uint}
	\int_{-\infty}^{+\infty}\dd{u}\chi_\mathcal{T}((u+s)/2)\chi_\mathcal{T}((u-s)/2) &= \int_{-\infty}^{+\infty} \dd{u}\frac{\mathcal{T}^4}{(\mathcal{T}^2+[u+s]^2)(\mathcal{T}^2+[u-s]^2)}\nonumber\\
	&=\frac{\pi}{2}\frac{\mathcal{T}^3}{ s ^2+\mathcal{T} ^2}\;.
	\end{align}
Thus we can write the generic response function with Lorentzian switching as
	\begin{equation}\label{eq:Response2}
	\mathcal{F}_F(\Omega,\mathcal{T})=\frac{\pi\mathcal{T}^3}{4}\int_{-\infty}^{+\infty}ds\; \frac{\mathcal{W}_F( s )}{ s ^2+\mathcal{T} ^2}\; e^{-i\Omega s }\;,
	\end{equation}
Let us now specialise for each of our examples.

\subsection{Linear scalar coupling}

\begin{figure}[htpb]
		\begin{center}
		\includegraphics[scale=0.75]{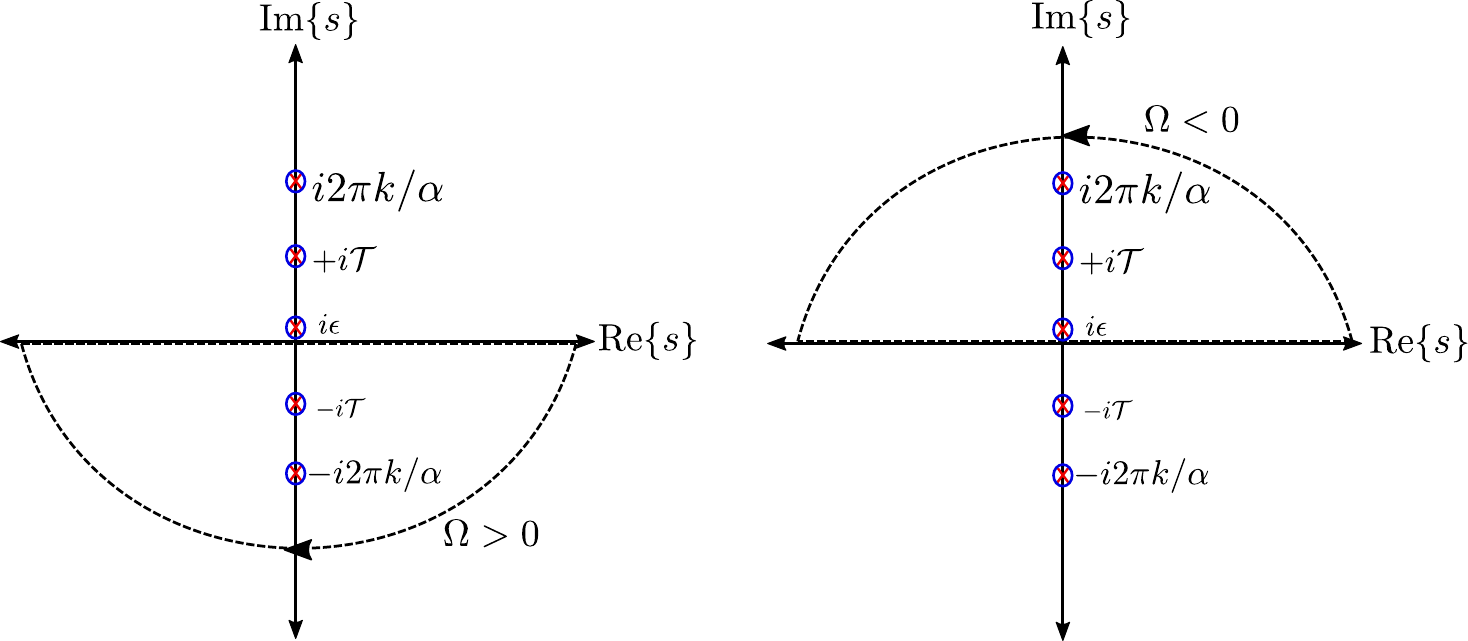}
	
		\caption[Pole Structure]{\label{fig:Poles}\small The pole structure of the response functions with Lorentzian regulator is independent of the coupling. The sign of $\Omega$ determines which contour we must choose.}
		\end{center}
	\end{figure}

In $d=4$ and using the relation 
	\begin{equation}
	\csch(z)^2=\sum\limits_{k=-\infty}^{\infty}(z-i\pi k)^{-2}\;,
	\end{equation}
the Wightman function of eq. (\ref{eq:SWight2}) takes the form 
	\begin{equation}
	\mathcal{W}_\phi( s )=-\frac{\alpha^2}{4\pi^2}\sum\limits_{k=-\infty}^{\infty}\frac{1}{(\alpha s -i\epsilon-2\pi ik)^2}\;.
	\end{equation}
Substituting this into eq.~(\ref{eq:Response2}), the response function in $d=4$ with a Lorentzian regulator takes the form
	\begin{equation}
	\mathcal{F}_\phi(\Omega,\mathcal{T})=-\frac{\alpha^2\mathcal{T}^3}{16\pi}\int_{-\infty}^{\infty}\dd{s}\sum\limits_{k=-\infty}^{\infty}\;\frac{e^{-i\Omega s }}{( \alpha s -i\epsilon-2\pi ik)^2( s ^2+\mathcal{T} ^2)}\;.
	\end{equation}
The general pole structure of this is shown in Figure~\ref{fig:Poles}.
Defining the dimensionless variables $x=\Omega/\alpha$, $\sigma=\alpha s $, and $y=\alpha\mathcal{T}$ we can split this into its parts 
	\begin{equation}
	\mathcal{F}(x,y)=I_0+\sum\limits_{k=1}^{\infty}I_k
	\end{equation}
where
	\begin{align}
	I_0&=-\frac{y^3}{16\pi}\int_{-\infty}^{+\infty}\dd{\sigma}\frac{e^{-ix\sigma}}{(\sigma^2+y^2)(\sigma-i\epsilon)^2}\;,\\
	I_k&=-\frac{y^3}{16\pi}\int_{-\infty}^{+\infty}\dd{\sigma}\frac{e^{-ix\sigma}}{(\sigma^2+y^2)}\left(\frac{1}{(\sigma-2\pi i k)^2}+\frac{1}{(\sigma+2\pi i k)^2}\right)\;,\; k\in \mathbb{Z}^+\;.
	\end{align}
We have taken the limit $\epsilon\rightarrow0$ in the second term as these remain finite for $k\neq0$. To evaluate $I_0$ we employ the residue theorem~\cite{Freitag2005}. For $x<0$ we must close the contour in the upper half plane which will pick up poles at $\sigma=iy$ and $\sigma=i\epsilon$. For $x>0$ we must close the contour in the lower half plane and will only pick up the pole at $\sigma=-iy$. Employing the formula 
	\begin{equation}\label{eq:res}
	\text{Res}(f,z_0)=\frac{1}{(n-1)!}\lim\limits_{z\rightarrow z_0}[(z-z_0)^nf^{(n-1)}(z)]
	\end{equation}
for the residue of a pole of order $n$~\cite{Freitag2005} we find after taking the limit $\epsilon\rightarrow0$
	\begin{equation}
	I_0=\frac{1}{16}\left(e^{-|x|y}+2|x|y\;\Theta(-x)\right) 
	\end{equation}
where $\Theta(x)$ is the Heaviside step function. Note that this is the response function for an inertial Unruh--DeWitt detector in Minkowski spacetime.

Considering now $I_k$, for $x<0$ we can close the contour in the upper half plane 
which will pick up a simple pole at $\sigma=iy$ and second order poles at $\sigma=2\pi i k$ for $k\in\mathbb{Z}^+$. Similarly for $x>0$ we must close the contour in the lower half plane and will pick up poles at $\sigma=-iy$ and $\sigma=-2\pi i k$ for $k\in\mathbb{Z}^+$. Let us fix $x>0$ and evaluate $I_k$ closing the contour in the lower half plane. Using the limit in eq.~(\ref{eq:res}) we obtain
	\begin{align}
	I_k 
	=&\frac{y^2e^{xy}}{16}\left(\frac{1}{(y+2\pi k)^2}+\frac{1}{(y-2\pi k)^2}\right)\nonumber\\
	&+\frac{y^2e^{-2\pi kx}}{64\pi^2}\left[\frac{1}{(k+\frac{y}{2\pi})^2}-\frac{1}{(k-\frac{y}{2\pi})^2}+2\pi x\left(\frac{1}{k+\frac{y}{2\pi}}-\frac{1}{k-\frac{y}{2\pi}}\right)\right]\;.
	\end{align}
The first terms come from the simple pole at $\sigma=-iy$ while the others are from the second order pole at $\sigma=-2\pi ik$. We can repeat a similar calculation for $x<0$ and all that will change are a few signs. 

Following ref.~\cite{Arias2018} we can use the definition of the Lerch--Hurwitz transcendent function~\cite{Ferreira2004} 
	\begin{equation}\label{eq:Lerch}
	\varPhi(z,n,a)=\sum\limits_{k=0}^{\infty}\frac{z^k}{(k+a)^n}\;
	\end{equation}
to find for general $x\in\mathbb{R}$
	\begin{align}
	\sum\limits_{k=1}^{\infty}I_k=&\frac{e^{-|x|y}}{16}\left(\frac{(y/2)^2}{\sin(y/2)^2}-1\right)\nonumber\\
	+&\frac{y^2e^{-2\pi|x|}}{64\pi^2}\left[\left(\varPhi(e^{-2\pi|x|},2,1+\frac{y}{2\pi})-\varPhi(e^{-2\pi|x|},2,1-\frac{y}{2\pi})\right)\right. \nonumber\\
	+&\left.2\pi|x|\left(\varPhi(e^{-2\pi|x|},1,1+\frac{y}{2\pi})-\varPhi(e^{-2\pi|x|},1,1-\frac{y}{2\pi})\right)\right]\;.
	\end{align}
Finally putting this all together the analytic expression for the response function in $d=4$ is 
	\begin{align}\label{eq:SResp}
	\mathcal{F}_\phi (x,y)=&\frac{1}{16}\left(2|x|y\;\Theta(-x)+\frac{(y/2)^2e^{-|x|y}}{\sin(y/2)^2}\right)\nonumber\\
	&+\frac{y^2e^{-2\pi|x|}}{64\pi^2}\left[\left(\varPhi(e^{-2\pi|x|},2,1+\frac{y}{2\pi})-\varPhi(e^{-2\pi|x|},2,1-\frac{y}{2\pi})\right)\right. \nonumber\\
	&+\left.2\pi|x|\left(\varPhi(e^{-2\pi|x|},1,1+\frac{y}{2\pi})-\varPhi(e^{-2\pi|x|},1,1-\frac{y}{2\pi})\right)\right]\;.
	\end{align}
which is the same as eq.~(B.14) in Arias et al.~\cite{Arias2018} except for a factor $1/2$, which we believe to be a simple omission. It also differs by a subtraction of $1/8$ which we believe was erroneously introduced in ref.~\cite{Arias2018} to ensure that in the limit $\mathcal{T}\rightarrow0$ we have $\mathcal{F}_F\rightarrow0$. This property, while desirable on physical grounds, is symptomatic of problems with finite time detectors. It has been argued that one must first take the limit $\mathcal{T}\rightarrow0$ before $\epsilon\rightarrow0$~\cite{Padmanabhan1996} for physically reasonable results, but the process is mathematically ill-defined. Since we are not interested in this limit we remain agnostic and just take $\epsilon\rightarrow0$. 
In the next sections we present results for quadratic and fermionic coupling.
\subsection{Quadratic coupling}
In the quadratic case using eq.~(\ref{eq:QExp}) and eq.~(\ref{eq:SWight2}) the vacuum correlator in $d=4$ is
		\begin{equation}\label{eq:QVCorr}
		\mathcal{W}_{\phi^2}(s)=\frac{\alpha^4}{128\pi^4}\frac{1}{\sinh^4[(\alpha s-i\epsilon)/2]}\;.
		\end{equation}
So using the same dimensionless variables the response function takes the form
	\begin{align}
	\mathcal{F}_{\phi^2}(x,y)&=\frac{\alpha^4\mathcal{T}^3}{512\pi^3}\int_{-\infty}^{+\infty}ds\; \frac{e^{-i\Omega s }}{ s ^2+\mathcal{T} ^2}\frac{1}{\sinh^4[(\alpha s-i\epsilon)/2]}\\
	&=\frac{\alpha^2y^3}{512\pi^3}\int_{-\infty}^{+\infty}d\sigma\; \frac{e^{-ix\sigma }}{ \sigma^2+y^2}\frac{1}{\sinh^4[(\sigma-i\epsilon)/2]}\;.
	\end{align}
This has the same pole structure (see Figure~\ref{fig:Poles}) as before except the poles at $\sigma=i\epsilon,2\pi ik$ for $k\in\mathbb{Z}$ are now 4th order. One can go through a similar procedure and extend the integration range to a contour through the complex plane and use the residue theorem to evaluate the integral, closing the contour in the lower (upper) half plane for $x>0$ ($x<0$).  However without using a series representation of the $\csch(z)^4$ function, this is a much more involved exercise. With the aid of {\sf Mathematica} we find 
%
	\begin{align}\label{eq:QResp}
		\mathcal{F}_{\phi^2}(x,y)&=\frac{y^4}{1536\pi^6\mathcal{T}^2}\left(3\pi^4e^{-|x|y}\csc(y/2)^4+4\pi^3e^{-2\pi|x|}\sum\limits_{n=1}^{4}\frac{P^{\phi^2}_n(x)\Delta\varPhi(x,y,n)}{(2\pi)^{n-1}}\right)\nonumber\\
		&+\frac{\Theta(-x)}{96\pi^2\mathcal{T}^2}\left(y^2P^{\phi^2}_1(x)+yP^{\phi^2}_3(x)\right)\;,
	\end{align}
where 
	\begin{equation}
		\Delta\varPhi(x,y,n)=\varPhi(e^{-2\pi|x|},n,1+\frac{y}{2\pi})-\varPhi(e^{-2\pi|x|},n,1-\frac{y}{2\pi})\;,
	\end{equation}
and 
	\begin{align}
	P^{\phi^2}_1&=|x|(1+x^2)\;,\;P^{\phi^2}_2=1+3x^2\;,\;P^{\phi^2}_3=6|x|\;,\;P^{\phi^2}_4=6\;.
	\end{align}
 Note that $P^{\phi^2}_{n+1}(x)=P^{\phi^2\prime}_{n}(x)$, except at $x=0$, and when $P^{\phi^2\prime}_{n}(x)$ has a discontinuity that contributes to the Heaviside term. It seems that this is the generic structure of the Unruh--Dewitt response functions with Lorentzian switching.
\subsection{Fermionic coupling}

The Fermionic case is much the same except we now have 6th order poles. From eq.~(\ref{eq:QExp}) and eq.~(\ref{eq:FExp}) the vacuum correlator takes the form
		\begin{equation}\label{eq:FVCorr}
			\mathcal{W}_\Psi(s)=-\frac{\alpha^6}{64\pi^4}\frac{1}{\sinh^6[(\alpha s-i\epsilon)/2]}\;.
		\end{equation}
Thus the fermionic response function is
	\begin{align}
		\mathcal{F}_{\Psi}(x,y) 
		&=-\frac{\alpha^4y^3}{256\pi^3}\int_{-\infty}^{+\infty}d\sigma\; \frac{e^{-ix\sigma }}{ \sigma^2+y^2}\frac{1}{\sinh^6[(\sigma-i\epsilon)/2]}\;,
		\end{align}
where again $x=\Omega/\alpha$, $\sigma=\alpha s $, and $y=\alpha\mathcal{T}$. Employing the residue theorem (see Figure~\ref{fig:Poles}) with the aid of {\sf Mathematica} we find 
%
	\begin{align}\label{eq:FResp}
		\mathcal{F}_{\Psi}(x,y)&=\frac{y^6}{3840\pi^8\mathcal{T}^4}\left(15\pi^6e^{-|x|y}\csc(y/2)^6+4\pi^5e^{-2\pi|x|}\sum\limits_{n=1}^{6}\frac{P^{\Psi}_n(x)\Delta\varPhi(x,y,n)}{(2\pi)^{n-1}}\right)\nonumber\\
		&+\frac{\Theta(-x)}{240\pi^2\mathcal{T}^4}\Big[P^{\Psi}_5(x)y+ P^{\Psi}_3(x)y^3 + P^{\Psi}_1(x)y^5\Big]\;,
	\end{align}
with $\Delta\varPhi(x,y,n)$ as before and 
	\begin{align}
	& P^{\Psi}_1=|x|(4+5x^2+x^4)\;,\;P^{\Psi}_2=(4+5x^2(3+x^2))\;,\;P^{\Psi}_3=10|x|(3+2x^2)\;,\nonumber\\
	& P^{\Psi}_4=30(1+2x^2)\;,\;P^{\Psi}_5=120|x|\;,\;P_6=120\;.
	\end{align}
 Again we have $P^{\Psi}_{n+1}(x)=P^{\Psi\prime}_{n}(x)$, except at $x=0$, and non-analytic $P^\Psi_n(x)$ terms appear with the Heaviside function. 

Some general remarks are now in order to help us understand these expressions. First, in each of these cases we can write  the difference $\mathcal{F}_F(x,y)-\mathcal{F}_F(-x,y)=-\Delta\mathcal{F}_F(x,y)$ in terms of a polynomial in $x$ and $y$ (this is simply the Heaviside term). It will be convenient to keep this separation. Another observation is that the dimensionality of the coupling constant $\lambda_F$ means that in $d=4$ dimensions only the scalar response function is dimensionless. For this reason we shall present results in terms of the dimensionless quantity $\tilde{\lambda}_F=\lambda_F/\xi^{\Delta_F}$. 

In the next section we demonstrate that these response functions,  eqs.~(\ref{eq:SResp}, \ref{eq:QResp}, \ref{eq:FResp}), reproduce the KMS thermal behaviour in the long time limit.

\section{Long time limit of the response functions}\label{App:LongTime}
In refs.~\cite{Fewster2016,Garay2016} it is shown that in the long time limit the detector's response is the Fourier transform of the Wightman function $\hat{\mathcal{W}}_\phi$, i.e.

	\begin{equation}
	\lim_{\mathcal{T}\rightarrow\infty}\frac{\mathcal{F}_\phi(\Omega,\mathcal{T})}{\mathcal{T}}=\frac{1}{2\pi}\lVert\hat{\chi}\rVert^2\;\hat{\mathcal{W}}_\phi(\Omega)\;,
	\end{equation}
where $\lVert\hat{\chi}\rVert^2=(1/\mathcal{T})\int_{\mathbb{R}}|\hat{\chi}(\omega)|^2\dd{\omega}=\pi^2/2$ for the Lorentzian switching function in eq.~(\ref{eq:Lorentz}). Since the Wightman function of an accelerated observer is a KMS state this indicates that the detector has thermalised. In this appendix we show that the response functions of detectors satisfy this limit.

To do this we shall employ the asymptotic form of the Lerch--Hurwitz transcendent derived by Ferreira and L\'opez~\cite{Ferreira2004}. Recalling the definition eq.~(\ref{eq:Lerch}),
	\begin{equation}
	\varPhi(z,n,a)=\sum\limits_{k=0}^{\infty}\frac{z^k}{(k+a)^n}\;.
	\end{equation}
Under the assumptions; $\{n,z\}\in\mathbb{C}$, $\arg(a)<\pi$, and $|z|<1$ for $\Re\{a\}\leq0$, or $z\in\mathbb{C}\backslash[1,\infty)$ for $\Re\{a\}>0$, this admits the expansion~\cite{Ferreira2004}
	\begin{equation}
	\varPhi(z,n,a)=\frac{1}{1-z}\frac{1}{a^s}+\sum\limits_{k=1}^{N-1}\frac{\text{Li}_{-k}(z)\;(n)_k}{k!}\;a^{-k-s}+O(a^{-N-s})\;.
	\end{equation}
Here $\text{Li}_{-k}(z)$ is the polylogarithm~\cite{Ferreira2004}  and $(n)_k=\Gamma(k+n)/\Gamma(n)$. All of the response functions depend on the difference
	\begin{equation}
	\Delta\varPhi(x,y,n)=\varPhi(e^{-2\pi|x|},n,1+\frac{y}{2\pi})-\varPhi(e^{-2\pi|x|},n,1-\frac{y}{2\pi})\;,
	\end{equation}
where $x=\Omega/\alpha$ and $y=\alpha\mathcal{T}$. To use the asymptotic expression we will assume we can let $y\rightarrow y+i\eta$, for infinitesimal $\eta$, and then freely take the limit $\eta\rightarrow0$ after employing the expansion. Under this condition
	\begin{equation}\label{eq:DPhiasymp}
	\Delta\varPhi(x,y,n)=\frac{1}{1-e^{-2\pi|x|}}\left(\frac{1}{(1+\frac{y}{2\pi})^n}-\frac{1}{(1-\frac{y}{2\pi})^n}\right)+O(y^{-1-n})\;.
	\end{equation} 
From eq.~(\ref{eq:SResp}) and eq.~(\ref{eq:DPhiasymp}) the linearly coupled scalar response function becomes
	\begin{align}
	\alpha\frac{1}{y}\mathcal{F}_\phi(x,y)&\sim\alpha\left[\frac{|x|}{8}\Theta(-x)+\frac{2\pi|x|e^{-2\pi|x|}}{1-e^{-2\pi|x|}}\frac{y}{64\pi^2}\left(\frac{1}{(1+\frac{y}{2\pi})}-\frac{1}{(1-\frac{y}{2\pi})}\right)\right]+O(y^{-1})\nonumber\\
	&\sim\frac{|\Omega|}{8}\left(\Theta(-\Omega)+\frac{1}{e^{+2\pi|\Omega|/\alpha}-1}\right)+O(y^{-1})\;.
	\end{align}
Therefore we have
	\begin{equation}\label{eq:SRLim}
	\frac{2\pi}{\lVert\chi\rVert^2}\lim\limits_{\mathcal{T}\rightarrow\infty}\mathcal{F}_\phi/\mathcal{T}=\frac{|\Omega|}{2\pi}\left(\Theta(-\Omega)+\frac{1}{e^{+2\pi|\Omega|/\alpha}-1}\right)
	\end{equation}
which is the Fourier transform of the Wightman function $\mathcal{W}_\phi$ (See Takagi~\cite{Takagi1986}) exactly as required.

We can repeat this calculation for the quadratically coupled scalar response function. From eq.~(\ref{eq:QResp}) and eq.~(\ref{eq:DPhiasymp}) we find
	\begin{align}
	\alpha\frac{1}{y}\mathcal{F}_{\phi^2}(x,y)&\sim\alpha^3\frac{|x|(1+x^2)}{96\pi^2}\left[\Theta(-x)+\frac{4\pi y\; e^{-2\pi|x|}}{1-e^{-2\pi|x|}}\left(\frac{1}{(1+\frac{y}{2\pi})}-\frac{1}{(1-\frac{y}{2\pi})}\right)\right]+O(y^{-1})\nonumber\\
	&\sim\frac{|\Omega|(\alpha^2+\Omega^2)}{96\pi^2}\left(\Theta(-\Omega)+\frac{1}{e^{2\pi|\Omega|/\alpha}-1}\right)+O(y^{-1})\;.
	\end{align}
Hence
	\begin{equation}\label{eq:QRLim}
	\frac{2\pi}{\lVert\chi\rVert^2}\lim\limits_{\mathcal{T}\rightarrow\infty}\mathcal{F}_{\phi^2}/\mathcal{T}=\frac{|\Omega|(\alpha^2+\Omega^2)}{24\pi^3}\left(\Theta(-\Omega)+\frac{1}{e^{2\pi|\Omega|/\alpha}-1}\right)\;,
	\end{equation}
which one can show exactly equals the Fourier transform of $\mathcal{W}_{\phi^2}(v)$ in eq.~(\ref{eq:QVCorr}). 

Finally taking the Fermionic response function eq.~(\ref{eq:FResp}) and the asymptotic form eq.~(\ref{eq:DPhiasymp}) we obtain
	\begin{align}
	\alpha\frac{1}{y}\mathcal{F}_{\Psi}(x,y)&\sim\alpha^5\frac{|x|(4+5x^2+x^4)}{240\pi^2}\left[\Theta(-x)+\frac{4\pi y\;e^{-2\pi|x|}}{1-e^{-2\pi|x|}}\left(\frac{1}{(1+\frac{y}{2\pi})}-\frac{1}{(1-\frac{y}{2\pi})}\right)\right]+O(y^{-1})\nonumber\\
	&\sim\frac{|\Omega|(4\alpha^4+5\alpha^2\Omega^2+\Omega^4)}{240\pi^2}\left(\Theta(-\Omega)+\frac{1}{e^{2\pi|\Omega|/\alpha}-1}\right)+O(y^{-1})\;.
	\end{align}
Thus
	\begin{equation}\label{eq:FRLim}
	\frac{2\pi}{\lVert\chi\rVert^2}\lim\limits_{\mathcal{T}\rightarrow\infty}\mathcal{F}_{\Psi}/\mathcal{T}=\frac{|\Omega|(4\alpha^4+5\alpha^2\Omega^2+\Omega^4)}{60\pi^3}\left(\Theta(-\Omega)+\frac{1}{e^{2\pi|\Omega|/\alpha}-1}\right)\;.
	\end{equation}
Again one can show this is the Fourier transform of $\mathcal{W}_{\Psi}(v)$ in eq.~(\ref{eq:FVCorr}). 

Hence we have shown, as expected, that the long time limit of the response functions, irrespective of the coupling, is a KMS state. And moreover the ratio of excitation to de-excitation probability is Boltzmann
	\begin{equation}
		\lim\limits_{\mathcal{T}\rightarrow\infty}\frac{\mathcal{F}_F(\Omega,\mathcal{T})}{\mathcal{F}_F(-\Omega,\mathcal{T})}=e^{-2\pi\Omega/\alpha}
	\end{equation} 
justifying the use of the vacuum as a thermal bath with Unruh temperature $T_U=\alpha/(2\pi)$.

\acknowledgments
This work was supported in part by the Natural Sciences and Engineering Research Council of Canada. F.G. was supported by Perimeter Institute through the Perimeter Scholars International Scholarship.

\bibliographystyle{JHEP}
\bibliography{FG_UQO_Refs} 

\end{document}